\documentclass[twocolumn,english,aps,superscriptaddress,prb]{revtex4-1}
\usepackage{graphicx} 
\usepackage{float}
\usepackage{tikz}
\usepackage{geometry}
\usepackage{amsmath}
\usepackage{amssymb}
\usepackage{physics}
\begin{document}
\title{Numerical investigation of quantum phases and phase transitions in a two-leg ladder of Rydberg atoms}

\author{Jose Soto Garcia}

\affiliation{Kavli Institute of Nanoscience, Delft University of Technology, Lorentzweg 1, 2628 CJ Delft, The Netherlands}
\author{Natalia Chepiga}
\affiliation{Kavli Institute of Nanoscience, Delft University of Technology, Lorentzweg 1, 2628 CJ Delft, The Netherlands}

\begin{abstract}
Experiments on chains of Rydberg atoms appear as a new playground to study quantum phase transitions in 1D. 
As a natural extension, we report a quantitative ground-state phase diagram of Rydberg atoms arranged in a two-leg ladder interacting via van der Waals potential.
We address this problem numerically, using the Density Matrix Renormalization Group (DMRG) algorithm. Our results suggest that,  quite remarkably, $\mathbb{Z}_k$ crystalline phases, with the exception of the checkerboard phase, appear in pairs characterized by the same pattern of occupied rungs but distinguishable by a spontaneously broken $\tilde{\mathbb{Z}}_2$ symmetry between the two legs of the ladder. Within each pair, the two phases are separated by a continuous transition in the Ising universality class, which  eventually fuses with the $\mathbb{Z}_k$ transition, whose nature depends on $k$. 
According to our results, the transition into the $\mathbb{Z}_2\otimes \tilde{\mathbb{Z}}_2$ phase changes its nature multiple of times including an Ashkin-Teller transition surprisingly stable over an extended interval, followed 
by the $\mathbb{Z}_4$-chiral transition and finally in a two step-process mediated melting via the floating phase. 
The transition into the $\mathbb{Z}_3$ phase with resonant states on the rungs belongs to the three-state Potts universality class at the commensurate point, to the $\mathbb{Z}_3$-chiral Huse-Fisher universality class away from it, and eventually it is through an intermediate floating phase. The Ising transition between $\mathbb{Z}_3$  and  $\mathbb{Z}_3\otimes \tilde{\mathbb{Z}}_2$ phases, entering the floating phase, opens new possibility to realize lattice supersymmetry in Rydberg quantum simulators.

\end{abstract}

\maketitle
\section{Introduction}
Understanding the nature of quantum phases and phase transitions is one of the central topics of modern condensed matter, high energy, and atomic physics\cite{sachdev1999quantum, carr2010understanding}. 
Over the last decades, the development of advanced numerical techniques such as the density matrix renormalization group (DMRG) and tensor network algorithms \cite{PhysRevLett.69.2863, ostlund1995thermodynamic, schollwock2011density, cDMRG} have led to numerous fascinating theoretical predictions. In this respect, tunable and controllable quantum simulators\cite{cirac2012goals, bloch2012quantum, blatt2012quantum, aspuru2012photonic, labuhn2016tunable, verresen2019stable, celi2020emerging, de2019observation} offer a unique opportunity to experimentally probe exotic quantum phenomena in low dimensions and stimulate further theoretical and numerical exploration. 
Among various platforms, neutral Rydberg atoms trapped with optical tweezers are particularly attractive for quantum simulations due to their long life time, strong interactions, and versatile geometries granted by an individual control over the atoms\cite{bernien2017probing,keesling2019quantum,zhang2024probing, de2019observation,scholl2021quantum, ebadi2021quantum}.

Neutral atoms in the ground state are coupled to the excited Rudberg state by a laser with Rabi frequency $\Omega$ a detuning $\Delta$. Atoms in the Rydberg state interact through van der Waals repulsion, which is known to lead to the phenomenon known as the Rydberg blockade: two atoms located within a certain distance $R_b$ cannot both be excited to the Rydberg state. The interplay between the laser detuning controlling the number of atoms in the Rydberg state, and the repulsive interactions between them gives rise to a rich variety of crystalline phases and critical phenomena even in 1D arrays\cite{keesling2019quantum,2019arXiv190802068R,chepiga2019floating,chepiga2021kibble,PhysRevResearch.4.043102, li2024uncovering}.

One of the fundamental problems  brought forward by experiments on  Rydberg atoms in 1D is the chiral melting\cite{huse1982domain} of the period-$\it p$ phases. One might expect that the transition out of phases with integer periodicity $\it p$ can be effectively  described in conformal field theory (CFT)\cite{difrancesco} by the corresponding minimal models: Ising for $p=2$, three-state Potts for $p=3$, Ashkin-Teller for $p=4$. However, due to long-range interactions between excited atoms, domain walls that bring particles closer to each other cost more energy than those sending particles further apart. This creates chiral perturbations that, according to Huse and Fisher\cite{huse1982domain,huse1984commensurate}, lead to a novel  non-conformal transition for $p=3$. Originally proposed in the 1980s in the context of 2+0D adsorbed monolayers, this exotic chiral transition remained an unsolved and intensely debated problem for several decades\cite{huse1982domain,schultz,haldane1983phase, huse1984commensurate,Den_Nijs,schreiner1994experimental,fendley2004competing}. Recently, large-scale DMRG simulations have fully resolved all critical regimes along  the boundary of the period-3 phase: a single conformal point in the 3-state Potts universality class emerges where the chiral perturbations vanish, followed by the Huse-Fisher chiral transitions up to the Lifshitz point, beyond which the chiral perturbations become too strong, leading to the emergence of a floating phase\cite{PBak_1982}---a critical Luttinger liquid phase characterized by quasi-long-range incommensurability\cite{chepiga2019floating,PhysRevResearch.4.043102,PhysRevA.98.023614,PhysRevB.99.094434,huang2019nonequilibrium, soto2024resolving}. Moreover, a similar sequence of quantum criticalities, including the conformal Ashkin-Teller point, the floating phase, and a novel $\mathbb{Z}_4$-chiral transition, has been numerically identified at the boundary of the period-4 phase\cite{chepiga2021kibble,PhysRevResearch.4.043102,PhysRevB.108.184425}. This rich theoretical picture is in full agreement with experimental results reported for 1D arrays of Rydberg atoms\cite{bernien2017probing}.  

The plethora of crystalline phases and an unprecedentedly rich critical behavior at the boundary of these phases motivate us to look beyond the simplest 1D chains of Rydberg atoms.  Experimentally, recent technological advancements have enabled remarkable control over the positioning of atoms, paving the way for the investigation of diverse lattice geometries beyond the single 1D chain\cite{labuhn2016tunable, de2019observation, ebadi2021quantum, scholl2021quantum, semeghini2021probing, chen2023continuous} reporting, in particular, a realization of the floating phase in the rectangular two-leg ladder\cite{zhang2024probing}.
Theoretical studies of Rydberg ladders, although still in their early stages, have already reported numerous exotic quantum phenomena that can potentially be realized in these systems. For instance, two-leg ladders of Rydberg-dressed atoms, conserving the number of particles, show complicated ground-state phase diagrams featuring the supersymmetric criticality and various types of Luttinger liquid phases\cite{Fromholz2022Phase, Tsitsishvili2022Phase}. Coupled Rydberg chains fine-tuned to their Ising
critical point open a route to realize a spin liquid\cite{verresen2021prediction, semeghini2021probing, slagle2022quantum}, while triangular ladders provide an alternative path for exotic dynamics\cite{2024arXiv241018913K}. A ladder with staggered detuning has been show to undergo the Ising transition induced by  the order-by-disorder mechanism\cite{sarkar2023quantum}. 
An effective blockade model of Rydberg atoms on a ladder has been shown to host two different charge density wave phases with spontaneously broken $\mathbb{Z}_3$ symmetry, Huse-Fisher chiral transition, and an extended Ashkin-Teller transition protected by integrability of the model\cite{eck2023critical}. 
The approximations made in each of these models, though made them compliant with the chosen analytical or numerical techniques, might significantly affect the original physics and even lead to contradictions\cite{eck2023critical,zhang2024probing}, as elaborated on later in the text. 
It is therefore rather surprising, that an accurate quantitative study of realistic models of Rydberg ladders with the van der Waals potential is still lacking. Our goal is to fill this gap by performing state-of-the art tensor network simulations. 


We explore in detail the phase diagram of a realistic model of Rydberg atoms on a two-leg ladder bearing in mind two motivations. Firstly, in quasi-1D, the competition between repulsive interactions along and between the legs enhances the symmetry of the problem. This increased symmetry might eventually lead to more complex synthetic phases, which cannot be realized within a single chain of Rydberg atoms, and to more exotic critical phenomena, including a wide variety of critical fusions. Secondly, we would like to resolve an open question on whether the chiral transitions observed in a single chain\cite{keesling2019quantum,PhysRevA.98.023614,chepiga2019floating,chepiga2021kibble,PhysRevResearch.4.043102} appear in a two-leg ladder of Rydberg atoms. Previous results are controversial in this respect: in a toy model with Rydberg blockade, only the chiral transition into $\mathbb{Z}_3$ has been found, while the transition into period-4 phase has been identified as either conformal with central charge $c=1$ or first order\cite{eck2023critical}. By contrast, an opposite scenario featuring the chiral transition into $\mathbb{Z}_4$ phase but lacking chiral transition into $\mathbb{Z}_3$ phase, has been predicted for a rectangular Rydberg ladder with the inter-atomic distance along the legs being two times smaller than the distance between the legs\cite{zhang2024probing}. In the present paper, we focus on a realistic model  with van der Waal interactions on a square ladder with equal inter-atomic distances along legs and rungs, and show a surprisingly rich phase diagram featuring an unusual pairwise appearance of the crystalline phases and (at least) two different chiral transitions. 


The rest of the paper is organized as follows: Section \ref{sec:overview} introduces the lattice model of the two-leg Rydberg ladder along with an overview of the main phases and phase transitions. In section \ref{sec:methods}, we explain the numerical methods and the protocols used to extract the critical exponents, along with the benchmark on the Ising transition into the checkerboard phase. Sections \ref{sec:z2z2} and \ref{sec:z3z2} present a detailed study of the quantum phase transitions out of $\mathbb{Z}_2$ and $\mathbb{Z}_2\otimes\tilde{\mathbb{Z}}_2$, and out of $\mathbb{Z}_3$ and $\mathbb{Z}_3\otimes\tilde{\mathbb{Z}}_2$ phases. Finally, we summarize our results and put them into perspective in Section \ref{sec:discussion}.

\section{Overview of the phase diagram}
\label{sec:overview}

We consider the following microscopic Hamiltonian defined in terms of hard-core-bosons:
\begin{equation}
          \frac{H}{\hbar} = \frac{\Omega}{2}\sum_i\left(d_i + d^\dagger_i\right) - \Delta\sum_i n_i + \sum_{i<j}V_{ij}n_in_j,
          \label{eq:ham}
\end{equation}
where $\Omega$ is the Rabi frequency and $\Delta$ is the laser detuning of the coherent laser with respect to the resonant frequency. The stability of the Rydberg state is controlled by $\Delta$, which acts as a chemical potential. Atoms in a Rydberg state interact via a repulsive van der Waals potential $V_{ij}$  that decreases with the distance $\abs{\vec r_{ij}}$ between the two excited atoms as $V_{ij}=V_0 \abs{\vec r_{ij}}^{-6}$, where $V_0$ is a constant that depends on the chosen Rydberg level. The van der Waals repulsion between Rydberg states induces a blockade effect, which prevents two atoms within a distance smaller than the blockade radius $R_b$, from being simultaneously excited to a Rydberg state. This is defined as  $V(R_b) = \Omega$, in other words $R_b/a= \left(V_0/\Omega\right)^{1/6}$, where $a$ is an interatomic distance.

For simplicity, we adopt the terminology of hard-core bosons, where atoms excited to a Rydberg state are referred to as ``occupied" ($n_i=1$), and atoms in the ground state are termed ``empty" ($n_i=0$). The creation (annihilation) operators are denoted by $d^\dagger_i$ ($d_i$), with the hard-core boson constraint forbidding multiple excitations, such that $d^\dagger_i\ket{1}_i = 0$.

\begin{figure}[h]
    \centering
    \includegraphics[width=\columnwidth]{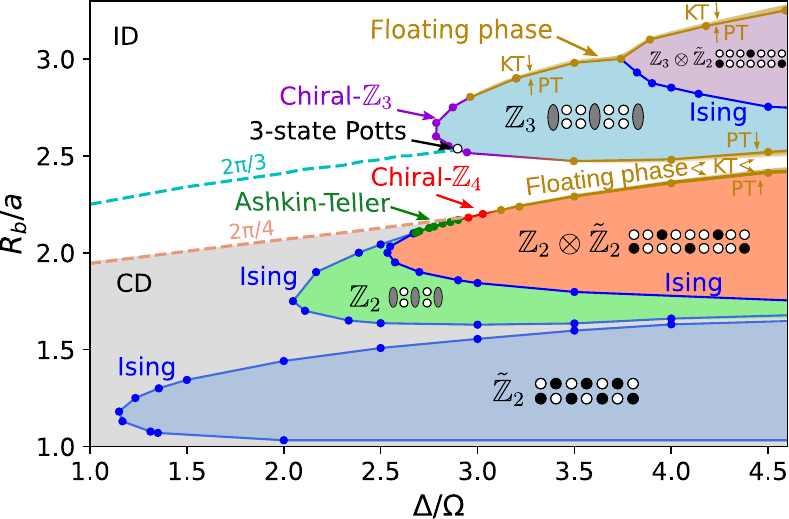}
    \caption{The ground-state phase diagram of a square two-leg ladder of Rydberg atoms for the blockade radius $R_b$ normalized to the interatomic distance $a$ against the laser detuning $\Delta$ normalized to the Rabi frequency $\Omega$. The blockade radius is defined as $R_b/a \equiv \left(V/\Omega\right)^{1/6}$. The diagram contains a disordered phase, five gaped phases with distinct crystalline structures and three narrow floating phases. The ordered phases are labeled according to their spontaneously broken symmetries. Floating phases are indicated by a dark yellow gradient line, reflecting the uncertainty in their precise extent. The sketches provide an intuitive visualization of one of the possible ground-states: filled circles state for the atoms in the excited Rydberg state, open circles for the atoms in the ground state, and gray ovals indicate a resonance state on a given rung. The disordered phase has a region with commensurate correlations (CD, gray) and incommensurate ones (ID, white). Dashed lines state for commensurate lines with $q=2\pi/4$ (orange) and $q=2\pi/3$ (cyan). There are four Ising transitions (blue). Transitions into $\mathbb{Z}_2\otimes\mathbb{\tilde Z}_2$ and into $\mathbb{Z}_3$ phases change its nature multiple times and, over a finite interval, fall within their corresponding chiral universality classes (red and purple).}
    \label{fig:phasediagram}
\end{figure}

Our main results are summarized in the phase diagram presented in Fig.\ref{fig:phasediagram}(a). The phase diagram features multiple lobes of crystalline phases, surrounded by a disordered phase. The disordered phase comprises two regions: one with commensurate density-density correlations (CD, gray area), and another with incommensurate correlations (ID, white area).  Within the incommensurate region, the dominant wave-vector changes continuously, and eventually takes commensurate values. For instance, the dashed turquoise line in Fig.\ref{fig:phasediagram}(a),\ with $q=2\pi/3$. Along this line, there are no chiral perturbations, and the transition to the corresponding $\mathbb{Z}_3$ phase is conformal in the 3-state Potts universality class.  

The lowest lobe corresponds to a checkerboard phase with one excited atom per rung, spontaneously breaking $\tilde{\mathbb{Z}}_2$ symmetry between the two legs. Remarkably, at larger values of $R_b/a$, all ordered lobes appear in pairs. Within each pair, the two phases can be distinguished by a spontaneously broken $\tilde{\mathbb{Z}}_2$ symmetry between the two legs.
The phase appearing below spontaneously breaks $\mathbb{Z}_k$ symmetry, forming a density wave phase along the chains. In the resonant phase, a single boson (excited Rydberg atom) occupies every $k$'s rung, and resonates between its two sites so that the local density profiles on both legs are identical. Naively, one would expect this phase to be fragile, but this resonant configuration appears to be a very stable one. The symmetry between the two legs is spontaneously broken in the lobe appearing just above, resulting in the phase with $\mathbb{Z}_k\otimes \tilde{\mathbb{Z}}_2$ broken symmetry.

\begin{figure}[h]
    \centering
    \includegraphics[width=\linewidth]{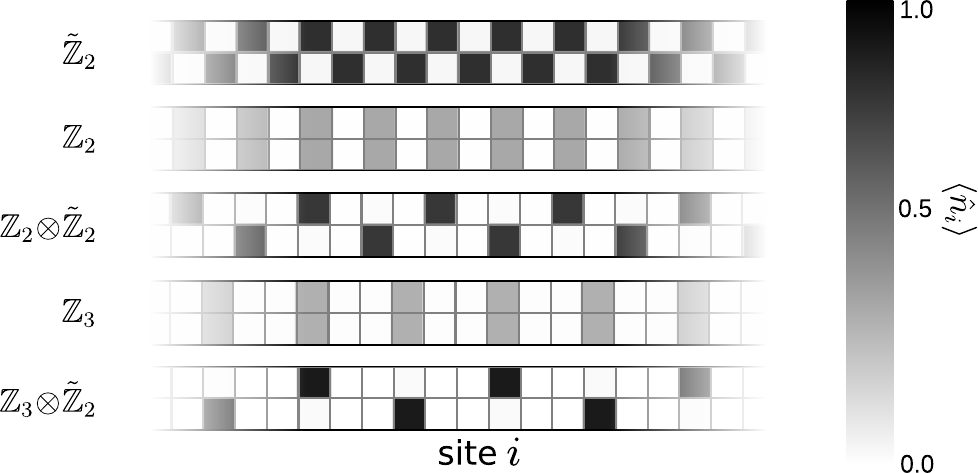}
    \caption{Typical density profiles of the gaped phases discussed in the main text. Each square corresponds to a lattice site, with its color indicating the local density according to the color bar. We show profiles  only for a central part of a long chain.}
    \label{fig:densitymap}
\end{figure}

In Fig.\ref{fig:densitymap}, we present the typical density profiles for each of the five ordered phases appearing in the phase diagram in Fig.\ref{fig:phasediagram}. For larger values of $R_b/a$ we also see the emergence of $\mathbb{Z}_4$ and $\mathbb{Z}_4\otimes \tilde{\mathbb{Z}}_2$ phases (see Appendix \ref{sec:z4} for examples of the density profiles) but due to growing computational costs we do not map out the accurate boundaries of these phases. In addition to the gaped ordered phases, we find a very thin floating phase surrounding the $\mathbb Z_2\otimes\tilde{\mathbb{Z}}_2$, $\mathbb Z_3$ and $Z_2\otimes\tilde{\mathbb{Z}}_2$ phases. The floating phase is critical Luttinger liquid phase a with quasi-long-range order and algebraically decaying incommensurate correlations. This phase emerges due to the strong chiral perturbation produced by the long-range interactions.

Let us now briefly list all phase transitions appearing in the phase diagram in Fig.\ref{fig:phasediagram}.
\begin{itemize}
    \item Disorder to $\tilde{\mathbb{Z}}_2$ (checkerboard): Ising
    \item Disorder to $\mathbb{Z}_2$ (period-2): Ising
    \item $\mathbb{Z}_2$ (period-2) to $\mathbb{Z}_2\otimes \tilde{\mathbb{Z}}_2$: Ising
    \item Disorder to $\mathbb{Z}_2 \otimes \tilde{\mathbb{Z}}_2$:
        \begin{itemize}
            \item Commensurate region: Ashkin-Teller
            \item Incommensurate region: Chiral/Floating Phase
        \end{itemize}
    \item Disorder to $\mathbb{Z}_3$
    \begin{itemize}
        \item Commensurate point: 3-state Potts
        \item Incommensurate region: Chiral/Floating Phase
    \end{itemize}
    \item $\mathbb{Z}_3$ to $\mathbb{Z}_3\otimes \tilde{\mathbb{Z}}_2$: Ising
    \item Disorder to $\mathbb{Z}_3 \otimes \tilde{\mathbb{Z}}_2$: Floating Phase
    \item In all cases, the transition from the disordered phase to the floating phase belongs to the Kosterlitz-Thouless\cite{Kosterlitz_Thouless} universality class, while the transition from the floating phase to the ordered gapped phase falls within the Pokrovsky-Talapov\cite{Pokrovsky_Talapov} universality class.
\end{itemize}

In the following sections, we discuss the details and present numerical evidences for all listed phase transitions.

\section{Methods}
\label{sec:methods}

\subsection{Ground-state calculations}

{\bf Approximation of algebraic interactions.} The numerical simulations have been performed with a state-of-the art density matrix renormalization group (DMRG) algorithm in the form of variational matrix product states (MPS)\cite{PhysRevLett.69.2863,schollwock2011density}.

\begin{figure}[h]
    \centering
    \includegraphics[width=\linewidth]{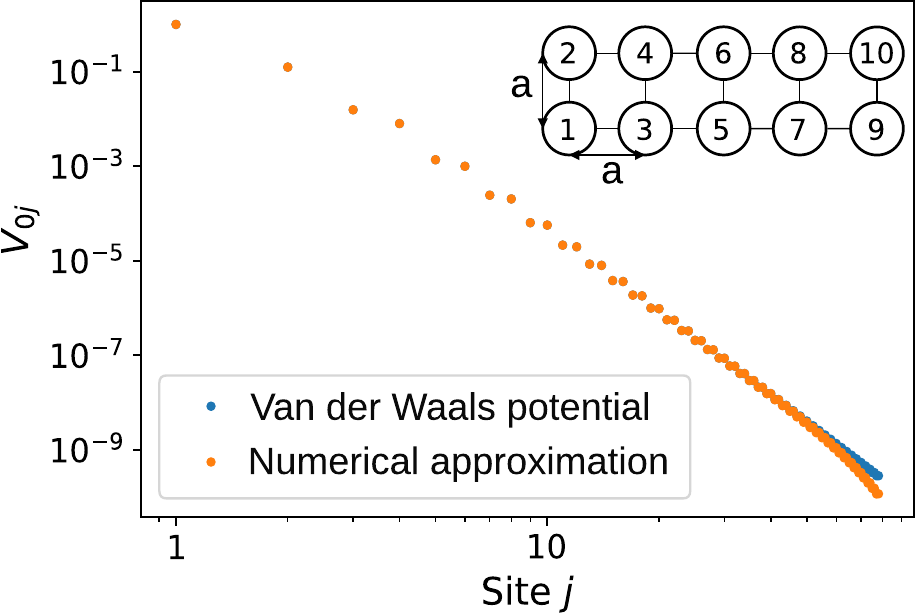}
    \caption{Comparison of the van der Waals repulsion between the first atom in the chain and atoms at subsequent positions along the linearized lattice (blue dots) with a tensor product approximation of the interaction (orange dots). For the tensor product approximation, the matrix representing the pairwise algebraic interactions of the ladder was approximated to a matrix product operator using a modified incremental singular value decomposition method \cite{lin2021isvd}. Note that the position on the lattice is not linearly proportional to the actual distance, as the atoms are arranged into a two-leg ladder.}
    \label{fig:vanderWaals}
\end{figure}

The matrix representing the pairwise algebraically decaying van der Waals interactions was approximated to a matrix product operator (MPO) using a modified incremental singular value decomposition method \cite{lin2021isvd}. The algorithm exploits the upper-triangular low-rank (UTLR) property of the pairwise interaction matrix. That is, the triangular portion of the matrix can approximated as a low-rank matrix. As a result, the process of constructing an MPO representation of the interaction matrix is reformulated as a matrix completion problem. We restricted the bond dimension of the MPO to $m = 11$, keeping the tolerance to be $10^{-9}$. As a result, the error on the approximation of the Hamiltonian was always smaller than $10^{-11}$. In Fig.\ref{fig:vanderWaals}, we show a comparison between  the van der Waals  potential scaling with the relative distance as $1/r^6$ and the numerical approximation we used in our numerical simulations.

{\bf Technical details.} We computed the ground states using the DMRG algorithm, keeping singular values above $10^{-8}$, and restricting the maximal bond dimension to $1200$. Convergence of the ground state was assumed when the relative energy difference between two successive sweeps (including an increase in the bond dimension)  was $\leq 10^{-11}$. The mentioned conditions allowed to reach convergence near criticality on systems with up to 700 rungs. 

{\bf Boundary conditions.} The number of rungs was always such that the ladder could accommodate the periodicity of the ordered phase. We applied fixed boundary conditions, but the way we fixed boundaries varied depending on whether the phase or the transition under consideration broke $\tilde{\mathbb{Z}}_2$ symmetry or not. When the symmetry between two legs was preserved, we applied a large and equal on-site potential on each site in the first and last rung. When the symmetry between two legs was broken, we applied a strong on-site potential on the top site of the first and last rungs in a ladder with a number of rungs $L=k n+1$, where $n$ is some integer and $k$ is periodicity of the $\mathbb{Z}_k$ phase. Deep inside the ordered phases, the effect of fixed boundaries disappears at the distances exceeding the correlation length, thus the ``wrong" choice of boundary conditions did not prevent us from seeing the correct ground-state of the bulk. In particular, and quite remarkably, favoring top or bottom sites at the edges does not destroy the resonance states deep in the bulk of the ladder.
In the critical regions, however, appropriate boundary conditions are essential to ensure the critical scaling and to reduce possible crossover effects.

\subsection{Extraction of the correlation length and the wave-vector}
Correlation length $\xi$ and wave-vector $q$ were extracted from  fitting the density-density correlation function to the Ornstein-Zernike function \cite{ornstein1914accidental, chepiga2021kibble}:
\begin{equation}
    C^{\textnormal{OZ}}_{ij} \propto \frac{e^{-\abs{i-j}/\xi}}{\sqrt{\abs{i-j}}}\cos\left(q\abs{i-j} + \varphi_0\right),
\end{equation}
where $\xi$ is the correlation length and $q$ is the wave vector. Both parameters, along with an initial phase $\varphi_0$, were treated as fitting parameters.

We extracted the connected correlation function $C_{ij}=\langle \hat O_i \hat O_j\rangle-\langle \hat O_i\rangle\langle \hat O_j\rangle$ numerically, where $\hat O_i$ are various density operators. In particular, in order to compute density-density correlations along  one leg of the ladder we used the operator $\hat t_i$ ($\hat b_i$), taking a value 1 if the top (bottom) site of the rung $i$ is occupied and 0 otherwise. In addition, we used symmetric and anti-symmetric combinations of these two operators in order to capture the resonating states:
\begin{equation}
\begin{aligned}
    \hat s_i = \hat t_i + \hat b_i \\
    \hat a_i = \abs{\hat t_i - \hat b_i}.
\end{aligned}
\label{eq:operator}
\end{equation}
In general, the extracted values of $\xi$ and $ q$ depend on the operator. 

\begin{figure}[h]
    \centering
   \includegraphics[width=\linewidth]{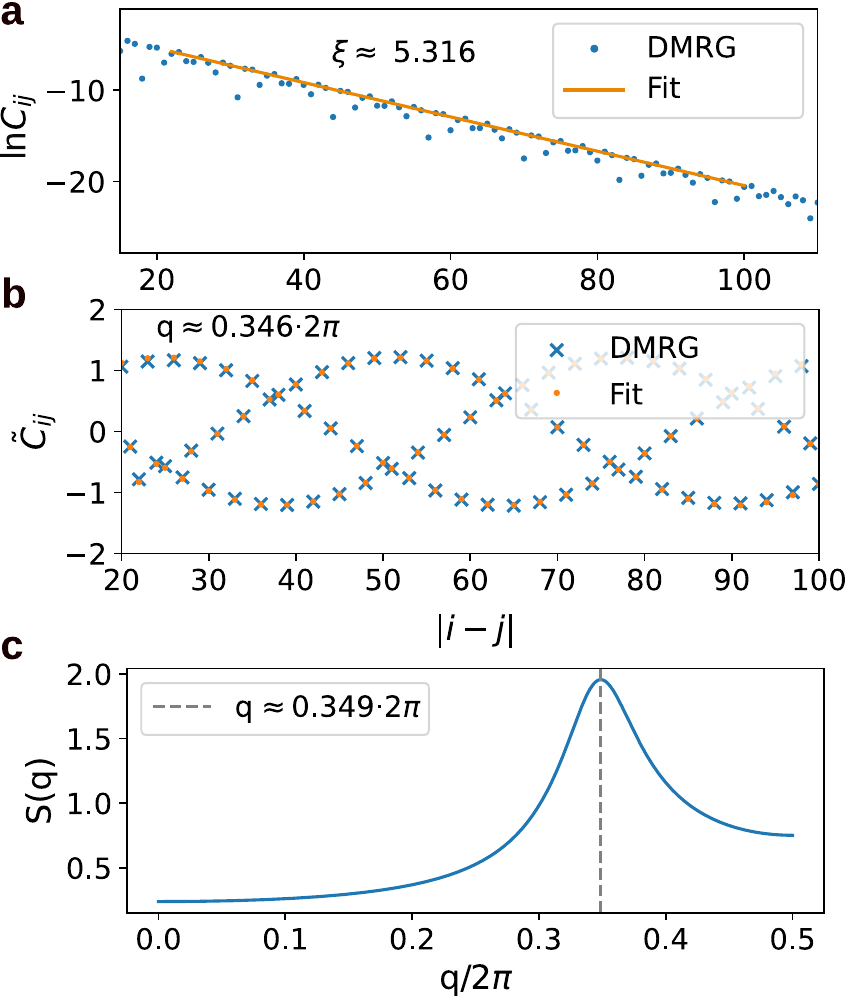}
    \caption{Example of extraction of the correlation length and wave-vector $ q$. (a) The correlation length is obtained by removing oscillations and fitting the correlation function on a semi-log plot. In this particular case, the operator for correlations is $\hat s$, defined in Eq.\ref{eq:operator} (b) The wave-vector $ q$ is extracted by removing the slope and fitting the oscillations to a cosine function. (c) The structure factor shows a peak at the same value of $q$ as the one obtained by fitting the oscillations. The data shown corresponds to $\Delta/\Omega = 3.0$ and $R_b/a = 2.4$.}
    \label{fig:enter-label}
\end{figure}

To extract $\xi$, oscillations were discarded and the slope of the decaying function was fitted in a semi-log scale with a linear function:
\begin{equation}
    \ln{C_{ij}} \approx c - x/\xi - 1/2\ln{|i-j|}.
\end{equation}
An example of such a fit is presented in Fig.\ref{fig:enter-label}(a), where correlations are calculated for operator $\hat s$.

In order to extract the wave-vector $q$,  we removed the main slope of the exponential decay from the correlation function $\tilde{C}=C_{ij} e^{ c - x/\xi - 1/2\ln{|i-j|}}$ and fit the rest with the oscillating cosine function:
\begin{equation}
    \tilde{C} \approx A\cos\left(q\abs{i-j} + \varphi_0\right),
\end{equation}
as demonstrated in the Fig.\ref{fig:enter-label}(b).

We noticed that, in the present case, $q$ can alternatively be extracted from the location of the peak of the operator-operator structure factor recently used to analyze incommensurability in experiments\cite{zhang2024probing}:
\begin{equation}
    S(q) \propto \sum_{i,i'}e^{jq(i-i')}\left<\hat s_i \hat \hat s_{i'}\right>,
\end{equation}
as presented in Fig.\ref{fig:enter-label}(c)\footnote{Note that instead of $\hat s_i$ one can use the operator $\hat a_i$ leading to the same values of the dominant wave-vector $q$.}. 

\subsection{Extraction of the central charge}
\begin{figure}[h]
    \centering
    \includegraphics[width=\linewidth]{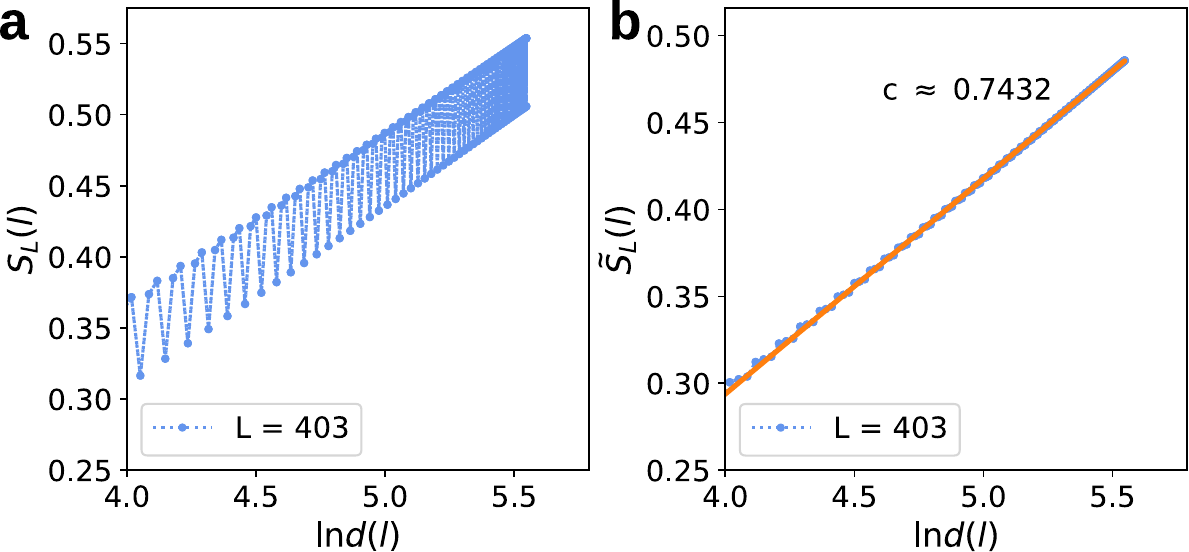}
    \caption{Scaling of the entanglement entropy $S$ with the conformal distance $d(l)$. (a) Oscillations in entanglement entropy are due to Friedel oscillations. (b) After removing the oscillations, the central charge can be extracted from a semi-log scaling. The data shown corresponds to $\Delta/\Omega$ = 2.78594 and $R_b/a = 2.6$.}
    \label{fig:cexample}
\end{figure}

According to conformal field theory (CFT), the entanglement entropy $S$ in a finite-size chain with open boundary conditions scales with the block size of the bipartite chain as\cite{calabrese2009entanglement, capponi2013quantum}: 
\begin{equation}
    S_{L}(l) = \frac{c}{6}\ln d(l) + s_1 + \ln {g}
\end{equation}
where $c$ is the central charge,  $d(l) = \frac{2L}{\pi}\sin(\frac{\pi l}{L})$ is the conformal distance, $ \ln g$ states for boundary entropy and $s_1$ is a non-uninversal constant. 
When the boundaries are fixed, Friedel oscillations break translation symmetry and contribute to the entanglement. One can remove the dominant contribution from Friedel oscillations:
\begin{equation}
    \tilde S_{L}(l) = S_L(l) + B\left(\left<\hat t_{l} \hat t_{l+1}\right> + \left<\hat b_{l} \hat b_{l+1} \right>\right),
\end{equation}
where $B$ takes some non-universal value. We present a typical scaling of entanglement entropy in Fig.\ref{fig:cexample}(a), shown before removing the Friedel oscillations, and in Fig.\ref{fig:cexample}(b), after removing them.

\subsection{Benchmark: Ising transition into the checkerboard phase}
\begin{figure}[t!]
    \centering
    \includegraphics[width=\linewidth]{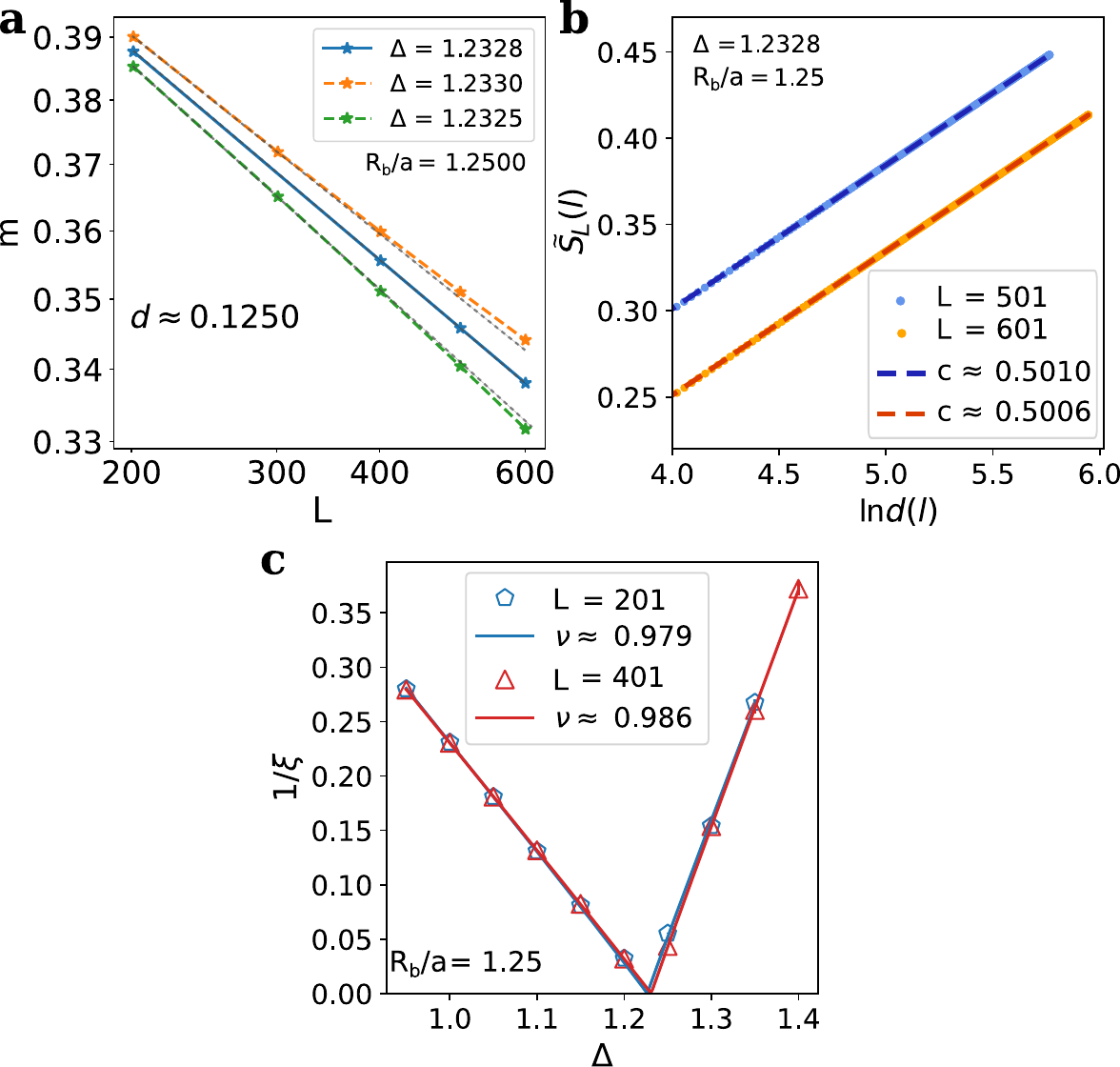}
    \caption{Numerical extraction of the critical exponents for the transition out of the $\tilde Z_2$ phase. a) Finite-size scaling of the order parameter $m = \hat a_{i}$ in the middle of a finite-size chain in a log-log scale. We identify the location of the critical point by the separatrix. To better highlight the curvature of the scaling curves, straight dashed lines are included as visual guides. The extracted scaling dimension $d$ is in excellent agreement with the CFT prediction for Ising transition $d = 1/8$. b) Scaling of the entanglement entropy $\tilde S_{L}$ with the conformal distance $d(l)$ after removing Friedel oscillations. Results for $L = 601$ where shifted by $-0.05$ for clarity. The slope agrees with the Ising central charge $c = 1/2$. c) Scaling of the inverse of the correlation length $1/\xi$ in the vicinity of the transition, extracted from the density-density correlations of the operator $\hat a$. The extracted critical exponent $\nu$ agrees with the theory prediction $\nu = 1$.}
    \label{fig:cb1criticalexponents}
\end{figure}

Let us use the transition into the checkerboard phase as a benchmark to establish the method we will apply throughout the paper for more complex transitions. The checkerboard phase spontaneously breaks $\tilde{\mathbb{Z}}_2$ symmetry, and therefore, it is expected to belong to the Ising universality class. As an example, in Fig.\ref{fig:cb1criticalexponents}, we present numerical results along an horizontal cut at $R_b/a = 1.25$. In order to locate the transition, we look at the scaling dimension of the order parameter $m \equiv \hat a_{L/2}$.
In the disordered phase, $m$ vanishes with the system size, while in the ordered phase, it eventually converges to a finite value. We thus associate the critical point with a separatrix that scales linearly with the system size in the log-log scale, as shown in Fig.\ref{fig:cb1criticalexponents}(a). We locate the critical point at $\Delta/\Omega = 1.2328$, $R_b/a = 1.25$. The slope of the separatrix gives the corresponding scaling dimension of the chosen operator. In the present case, the numerically extracted scaling dimension $d\approx0.125$ is in excellent agreement with the CFT prediction for the Ising transition\cite{difrancesco} $d = 1/8$. At this critical point, we extract the central charge by fitting the entanglement entropy, as shown in Fig.\ref{fig:cb1criticalexponents}(b). Our results are in excellent agreement with the CFT prediction $c=1/2$.

Finally, we extract the correlation length critical exponent $\nu$. To achieve this, we measure the correlation length on both sides of the transition and plot its inverse, as shown in Fig.\ref{fig:cb1criticalexponents}(c). We fit these data points on both sides of the transitions, assuming a single critical point and identical critical exponents in the ordered and disordered phases ($\nu=\nu^\prime$), while allowing for different non-universal pre-factors in each phase. The extracted critical exponent agrees within $2\%$ of the CFT value $\nu=1$.

\section{Transitions into $\mathbb{Z}_2$ and $\mathbb{Z}_2\otimes\tilde{\mathbb{Z}}_2$ phases}
\label{sec:z2z2}
In this section, we focus on the pair of phases in which every other rung of the ladder is occupied by a boson. As mentioned earlier, there are two possible configurations: either a boson resonates between the top and bottom sites of the occupied rung, or it occupies one of the two sites, spontaneously breaking the symmetry between the two legs. The two symmetries, $\mathbb{Z}_2$ and $\tilde{\mathbb{Z}}_2$, can be broken successively through two Ising transitions. From this perspective, the $\mathbb{Z}_2$ phase appears as an intermediate phase between the disordered and $\mathbb{Z}_2 \otimes \tilde{\mathbb{Z}}_2$ phases. 

\begin{figure}[b]
    \centering
    \includegraphics[width=\linewidth]{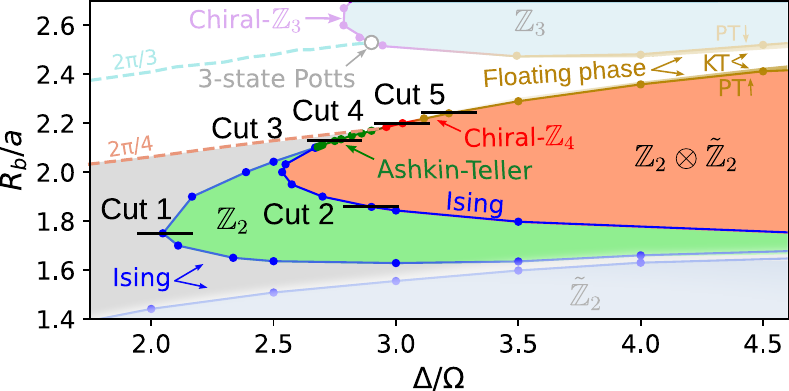}
    \caption{Part of the phase diagram presented in Fig.\ref{fig:phasediagram}, zoomed around a pair of phases with spontaneously broken $\mathbb{Z}_2$  symmetry, where, up to quantum fluctuations, every other rung of the ladder is occupied by a boson. The transition between the commensurate disordered phase (gray) and $\mathbb{Z}_2\otimes \tilde{\mathbb{Z}}_2$ phase occurs either  through a pair of Ising transitions (blue), breaking $\mathbb{Z}_2$ and $\tilde{\mathbb{Z}}_2$ symmetries sequentially, or through an Ashkin-Teller transition (green). In the region where the disordered phase is incommensurate (white), the transition to the  $\mathbb{Z}_2\otimes \tilde{\mathbb{Z}}_2$ phase is via a direct non-conformal $\mathbb{Z}_4$-chiral transition or through a floating phase---a narrow critical phase separated from the disordered and ordered phases by the Kosterlitz-Thouless (KT) and Pokrovsky-Talapov (PT) transitions, respectively. The location of the floating phase is indicated by a thick dark yellow gradient line, reflecting the uncertainty in the precise location of the KT transition.}
    \label{fig:z2sketch}
\end{figure}

Interestingly, at larger values of $R_b/a$, the two symmetries are broken simultaneously, which leads to a unique sequence of exotic transitions: an interval of the Ashkin-Teller conformal transition is followed by an interval of the non-conformal chiral transition, and further along, a narrow intermediate floating phase emerges between the disordered and $\mathbb{Z}_2 \otimes \tilde{\mathbb{Z}}_2$ phases. We provide numerical evidences for all these transitions, focusing in particular on the five cuts indicated in Fig.\ref{fig:z2sketch}.

\subsection{Ising transitions}

Let us first discuss the pair of Ising transitions. 
The first Ising transition between the disordered and $\mathbb{Z}_2$ phase spontaneously breaks translation symmetry by two rungs. We therefore define the order parameter as an amplitude of oscillations of total density on the rung $m=|\hat s_i- \hat s_{i+1}|$. In Fig.\ref{fig:Z2andZ2Z2ising}(a), we perform a finite-size scaling of this operator to locate the critical line and to extract the scaling dimension $d$. At the identified critical point, we extract the central charge by fitting the scaling of the entanglement entropy, as shown in Fig.\ref{fig:Z2andZ2Z2ising}(b). Finally we extract the correlation length critical exponent $\nu$ by fitting the inverse of the correlation length on both sides of the transition.
Numerically extracted values are in excellent agreement with the CFT predictions for an Ising transition\cite{difrancesco} $d=1/8$, $c=1/2$ and $\nu=1$.

\begin{widetext}
   \centering
   
\begin{figure}[h!]
    \includegraphics[width=0.8\textwidth]{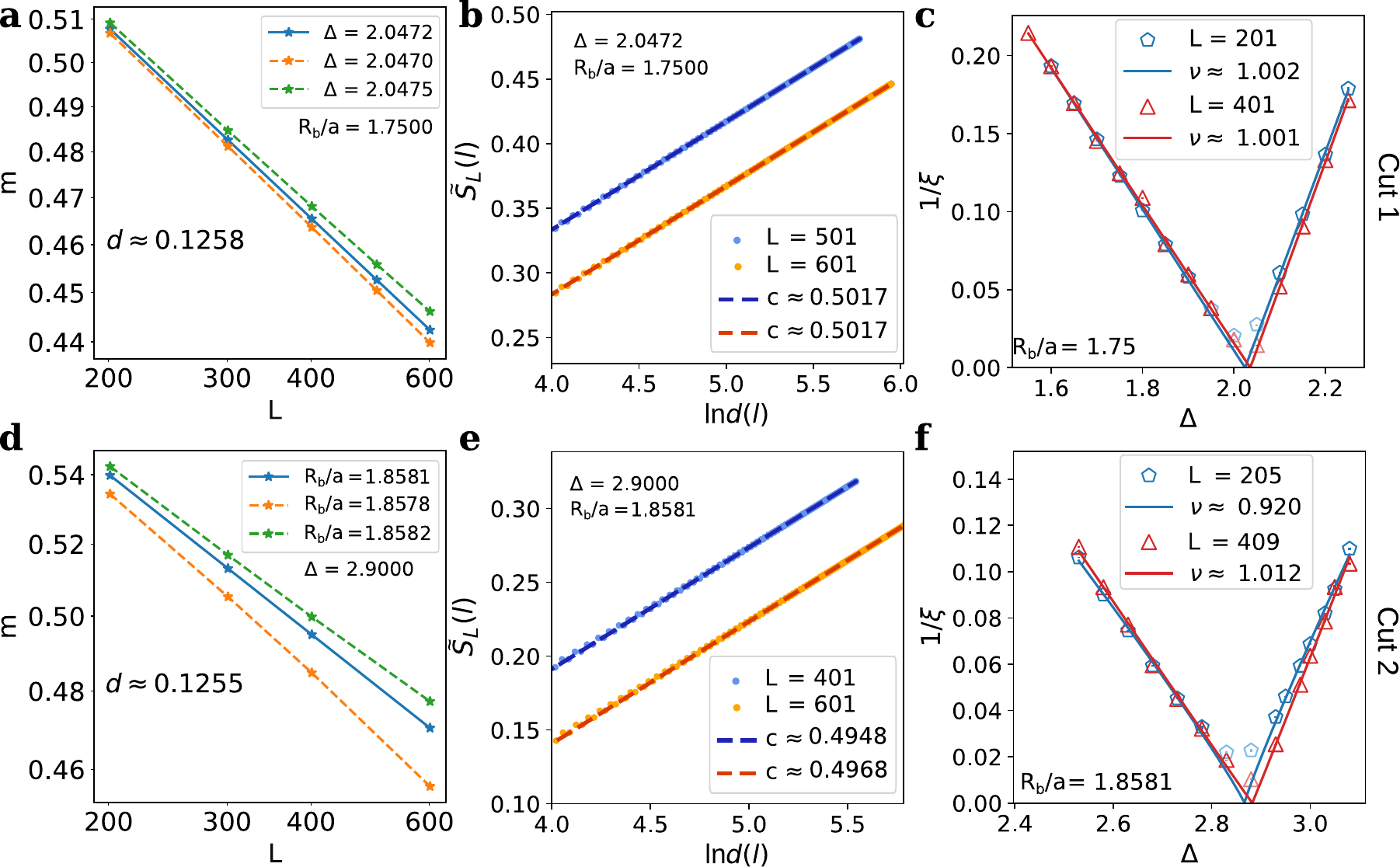}
    \caption{ Numerical evidence of the Ising transitions between (a)-(c) the disordered and $\mathbb{Z}_2$ ordered phases; and (d)-(f) the $\mathbb{Z}_2$ and $\mathbb{Z}_2\otimes \tilde{\mathbb{Z}}_2$ ordered phases. (a),(d) Finite-size scaling of the order parameter (a) $m=|s_i-s_{i+1}|$ and (b) $m=\max(\hat a_i, \hat a_{i+1})$ in a log-log scale. We identify the location of the critical point with the separatrix, its slope corresponds to the scaling dimension $d$ of the order parameter. Numerical results are in excellent agreement with CFT predictions for Ising transition $d=1/8$. (b),(e) Finite-size scaling of the reduced entanglement entropy. Numerically extracted values of the central charges are in excellent agreement with Ising transition $c=1/2$. Results for $L = 601$ were shifted by $-0.05$ for clarity. (c),(f) Scaling of the inverse of the correlation length $1/\xi$ on both sides of the transition extracted from the density-density correlations of the operator $\hat a$. In both cases the critical exponent $\nu$ agree within $1\%$ with the CFT prediction $\nu=1$. Small shift in the location of the critical point is a typical finite-size effect\cite{chepiga2019floating,chepiga2021kibble,PhysRevResearch.4.043102}. }
    \label{fig:Z2andZ2Z2ising}
\end{figure}

\end{widetext}

We perform a similar analysis for the second Ising transition---between $\mathbb{Z}_2$ and $\mathbb{Z}_2\otimes \tilde{\mathbb{Z}}_2$ phases. This transition is characterized by a spontaneously broken symmetry between  upper and lower legs. Therefore, we identify the order parameter as the difference in local density taken on one rung  $m=\max(\hat a_i, \hat a_{i+1})$. The maximum over two consecutive rungs is taken to select an occupied rung. 
As in the previous case, we extract the scaling dimension of the order parameter, the central charge, and the correlation length critical exponent. The extracted results are in excellent agreement with Ising predictions, and the examples of the scaling are presented in Fig.\ref{fig:Z2andZ2Z2ising}(d)-(f).

\subsection{Disorder to $\mathbb{Z}_2 \otimes \tilde{\mathbb{Z}}_2$ transition}

Eventually, the two Ising transitions discussed above approach each other and fuse into a multicritical Ashkin-Teller point. Beyond this point, both symmetries, $\mathbb{Z}_2$ and $\tilde{\mathbb{Z}}_2$, are broken simultaneously, and the nature of the transition undergoes multiple changes. As long as the disordered phase adjacent to the $\mathbb{Z}_2 \otimes \tilde{\mathbb{Z}}_2$ is commensurate (gray area in Fig.\ref{fig:z2sketch}) the transition remains conformal in the Ashkin-Teller universality class. However, beyond a certain threshold, the disordered phase exhibits short-range incommensurability. The line where the wave-vector $q$ starts to deviate from its commensurate value $q=2\pi/4$ is indicated in Fig.\ref{fig:z2sketch} with an orange dashed line. Our results are consistent with the Ashkin-Teller conformal transition below this line, and the $\mathbb{Z}_4$-chiral transition is eventually replaced by the floating phase above it. 

\begin{figure}[t]
    \includegraphics[width=0.45\textwidth]{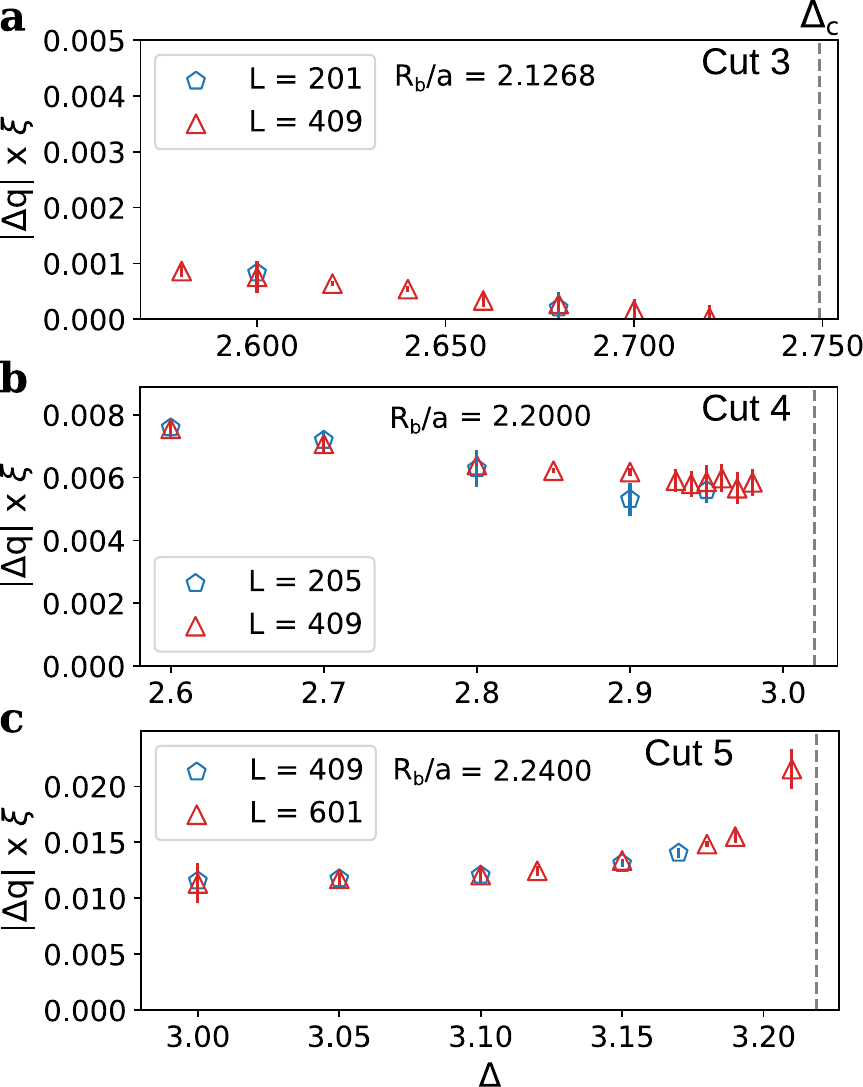}
    \caption{ Scaling of the product $\Delta q\times\xi$ upon approaching (a) Ashkin-Teller; (b) $\mathbb{Z}_4$-chiral; and (c) the Kosterlitz-Thouless transitions. The correlation length $\xi$ was extracted from the density-density correlations of the operator $\hat a$. The vertical dashed line indicates the location of the critical point. The scaling is performed along three horizontal cuts indicated in Fig.\ref{fig:z2sketch}. The error bars represent the $95\%$ confidence interval for the point's location, obtained via error propagation, accounting for the uncertainties in the fitted parameters and their correlations.}
    \label{fig:Z2Z2chiral}
\end{figure}

In order to distinguish the chiral transition from the conformal Ashkin-Teller, we use Huse-Fisher criterion\cite{huse1982domain} $\Delta q\times\xi$, where $\Delta q$ measures the distance of the incommensurate wave-vector $q$ to its commensurate value. In this particular case, $\Delta q=|q-2\pi/4|$. The product $\Delta q\times\xi$ is expected to vanish upon approaching the conformal transition, to take a finite value at the chiral transition, and to diverge at the Kosterlitz-Thouless transition into the floating phase. In Fig.\ref{fig:Z2Z2chiral}, we present the scaling of this product along the three cuts discussed in Fig.\ref{fig:z2sketch}, which go through different critical regimes. In Appendix \ref{sect:operatorcorrelation}, we provide further details on the operators that we used to extract $q$ and $\xi$ across various transitions.

The appearance of an extended interval of the Ashkin-Teller conformal transition is rather unusual. Previous studies of one-dimensional Rydberg arrays\cite{chepiga2021kibble,PhysRevResearch.4.043102} and ladders\cite{zhang2024probing} featured a single Ashkin-Teller point appearing at the intersection of the critical and commensurate lines. In multi-component Rydberg array\cite{chepiga2024tunable} and quantum loop ladder\cite{la2024mathbb}, where, similar to the present case, two Ising transitions fuse into a direct $\mathbb{Z}_2\otimes\mathbb{\tilde Z}_2$ transition, only the multicritical point was in the Ashkin-Teller universality class, followed by the chiral transition immediately after. Here, the extended interval of the Ashkin-Teller transition arises due to the fact that two Ising transitions fuse before the incommensurability develops. 

\begin{figure}[h]
    \centering
    \includegraphics[width=\columnwidth]{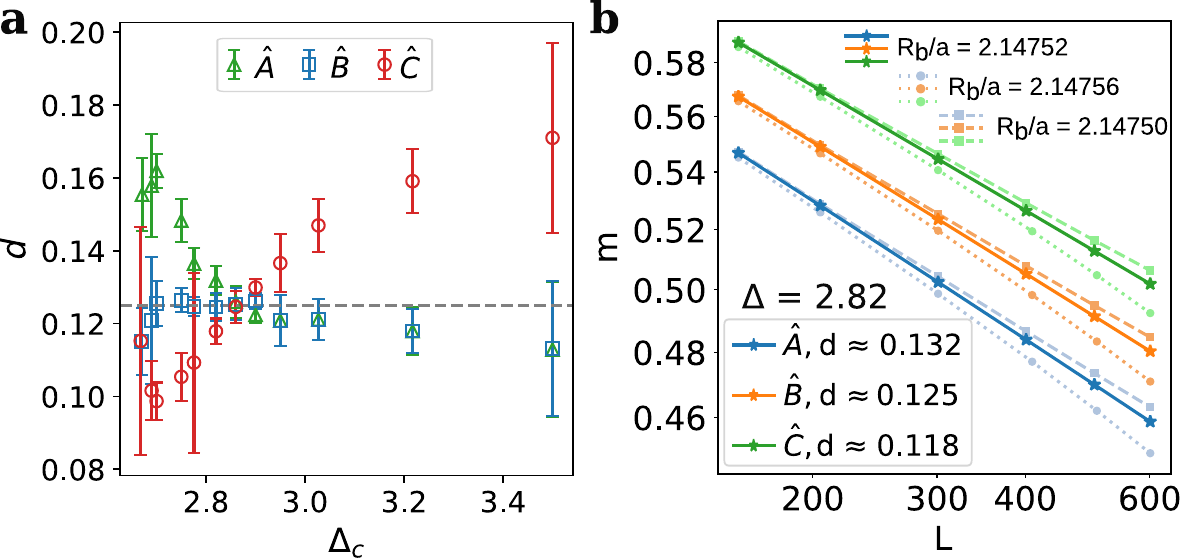}
    \caption{Scaling dimension $d$ calculated from the finite-size scaling of three different order parameters $\hat A$, $\hat B$ and $\hat C$ defined by Eq.\ref{eq:fssoperator} and measured at the centre of the chain. (b) Example of the extraction of the critical exponent $d_{\hat A,\hat B,\hat C}$ from the finite-size scaling of the order parameters. Error bars correspond to the $95\%$ confidence interval for the fit, estimated as $\pm 1.96$ times the standard error of the predicted values, obtained from the covariance matrix of the fit parameters.} 
    \label{fig:fsscuts}
\end{figure}

To further support the evidence of an extended conformal Ashkin-Teller transition, we present in Fig.\ref{fig:fsscuts} the scaling behavior of three order parameters, defined as follows:
\begin{equation}
\begin{aligned}
    \hat A &= \textnormal{max}\left(\hat a_{L/2}, \hat a_{L/2+1}  \right) \\
    \hat B &= \textnormal{max}\left(\hat t_{L/2-1}, \hat t_{L/2}, \hat t_{L/2+1}, \hat t_{L/2+2}\right) \\
    &\quad - \textnormal{min}\left(\hat t_{L/2-1}, \hat t_{L/2}, \hat t_{L/2+1}, \hat t_{L/2+2}\right) \\
    \hat C &= \abs{\hat s_{L/2} - \hat s_{L/2+1}}.
\end{aligned}
\label{eq:fssoperator}
\end{equation}
Operator $\hat A$ measures a spontaneously broken symmetry between the two chains. It is the same operator used to locate the Ising transition in Fig.\ref{fig:Z2andZ2Z2ising}(b). Operator $\hat B$ measures the amplitude of the local density over four consecutive sites along one chosen leg (here - top). This operator measures the spontaneously broken $\mathbb{Z}_4$ symmetry---the result of simultaneously breaking $\mathbb{Z}_2$ and  $\tilde{\mathrm{Z}}_2$. Consequently, we expect this operator to have a scaling dimension $d_{\hat B}=1/8$, corresponding to the Ashkin-Teller order parameter ($\beta/\nu$ in Baxter's notations for the eight-vertex model \cite{Baxter:1982zz}). This operator corresponds to $s_j$ in Eck and Fendley\cite{eck2023critical}, where the emergence of the Ashkin-Teller transition over an extended interval was preserved by integrability of their model. Operator $\hat C$ measures the amplitude of the total rung density, and therefore, reflects the order parameter associated with the spontaneously broken  $\mathbb{Z}_2$ translation symmetry. In Baxter's solution for the eight-vertex model, its scaling dimension corresponds to $\beta_e/\nu$\cite{Baxter:1982zz,PhysRevB.107.L081106}. This value is not universal but varies along the critical line, as we also observe in Fig.\ref{fig:fsscuts}.  Returning to operator $\hat A$, its meaning is not clear in the Baxter's theory for the eight-vertex model. However, it is reasonable to expect that if the order parameter associated with $\mathbb{Z}_2$ symmetry is non-universal, then its counterpart---the order parameter of $\tilde{\mathbb{Z}}_2$---will also be non-universal, as we see in Fig.\ref{fig:fsscuts}. In Eck and Fendley\cite{eck2023critical}, operator $\hat A$ is referred to as $(n_j^++n_j^-)^-$, and its scaling dimension was also reported to vary along the critical line, consistent with our findings. In Fig.\ref{fig:fsscuts}(b), we provide an example of how the scaling dimension for all three operators was extracted.

Based on the extracted scaling dimension $d_{\hat B}$, we identify the interval of the Ashkin-Teller transition as $2.7\lesssim \Delta\lesssim 2.95$. Within this interval, the numerically extracted central charge agrees with the CFT prediction $c=1$ within a $5\%$ (see Appendix \ref{sec:centralcharge}). Beyond $\Delta\gtrsim 2.95$, the transition is no longer conformal, and $d_{\hat B}\neq 1/8$.  Before that, at $2.67\lesssim \Delta\lesssim 2.7$, we see that $d_{\hat B}$ deviates from this value, probably due to proximity to the multicritical point and possible crossover region. Note that $d_{\hat C}$ takes values noticeably different from the value $1/8$ that it takes at the Ising transition between disordered and $\mathbb{Z}_2$ phases. This observation, combined with a large central charge in this region, allows us to confidently exclude the scenario in which the two Ising transitions continue up to $\Delta_C\approx 2.95$, with a single Ashkin-Teller point. Moreover, starting from $\Delta=2.75$, the locations of the critical points identified associated with a separatrix for each of the three order parameters $\hat A$, $\hat B$ and $\hat C$ is different by no more than $O(10^{-5})$.


\section{Transitions into $\mathbb{Z}_3$ and $\mathbb{Z}_3\otimes\tilde{\mathbb{Z}}_2$ phases}
\label{sec:z3z2}
Now, let us focus on the pair of phases where every third rung is occupied, illustrated in Fig.\ref{fig:z3z3z2sketch}. The bi-lobe structure consists of two phases: $\mathbb{Z}_3$ and $\mathbb{Z}_3\otimes\tilde{\mathbb{Z}}_2$, where, as in the previous case, $\tilde{\mathbb{Z}}_2$ refers to a spontaneously broken symmetry between the two legs. In the limit of strong chiral perturbations, an intermediate floating phase is expected to emerge, separating each of the $\mathbb{Z}_3$ phases from the disorder phase. Since the floating phase is an incommensurate Luttinger liquid phase with an emergent U(1) symmetry, we expect it to be separated from the crystalline $\mathbb{Z}_3$ phases by a commensurate-incommensurate Pokrovsky-Talapov\cite{Pokrovsky_Talapov} transition and from the disordered phase by the Kosterlitz-Thouless\cite{Kosterlitz_Thouless} transition transparent for incommensurability. As we shall see below, the floating phase appearing at the boundary of the $\mathbb{Z}_3$ lobes is very narrow, and we detect its appearance by a clear signature of the Kosterlitz-Thouless transition on one side of it. 

\subsection{Transition between the disordered and $\mathbb{Z}_3$ phases}
\begin{figure}[b]
    \centering
    \includegraphics[width=\linewidth]{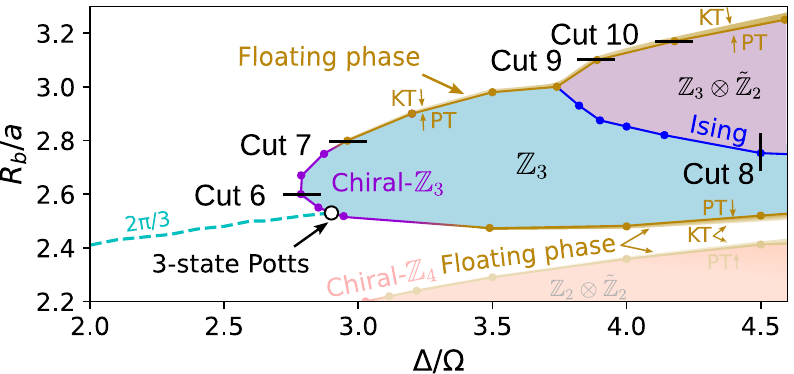}
    \caption{Part of the phase diagram presented in Fig.\ref{fig:phasediagram}, zoomed around a pair of phases with spontaneously broken $\mathbb{Z}_3$  symmetry, which, up to quantum fluctuations, corresponds to every third rung of the ladder being occupied. At the intersection of the commensurate line (dashed blue) with $q=2\pi/3$, the transition into the $\mathbb{Z}_3$ phase is in the three-state Potts universality class (open circle). Away from this point, but not too far, the transition falls within the Huse-Fisher $\mathbb{Z}_3$-chiral universality class, and further away, the transition is mediated by a narrow floating phase (gradient dark yellow) separated from the disordered phase by a Kosterlitz-Thouless (KT)  and from the ordered phase by a Pokrovsky-Talapov (PT) transitions.  The transition between the $\mathbb{Z}_3$ and $\mathbb{Z}_3\otimes \tilde{\mathbb{Z}}_2$ phases is Ising (blue). The transition between the disordered and $\mathbb{Z}_3\otimes \tilde{\mathbb{Z}}_2$ phases is always through a floating phase. The gradient shading in the floating phase reflects the uncertainty in its precise boundaries.}
    \label{fig:z3z3z2sketch}
\end{figure}

When the symmetry between two legs is preserved, the nature of the transition into $\mathbb{Z}_3$ phase is essentially identical to the one observed in the 1D chain of Rydberg atoms\cite{keesling2019quantum, chepiga2019floating}. Specifically, there is a single point where the commensurate line with $q = 2\pi/3$ in the disordered phase intersects the critical line, resulting in a conformal transition within the three-state Potts universality class, as illustrated in Fig.\ref{fig:z3z3z2sketch}. Away from this point, but not too far, the transition is chiral in the Huse-Fisher universality class, and further away, the transition is a two-step process via Kosterlitz-Thouless and Pokrovsky-Talapov transitions, with an extremely narrow floating phase in between.

\begin{figure}[h]
    \centering
    \includegraphics[width=0.9\linewidth]{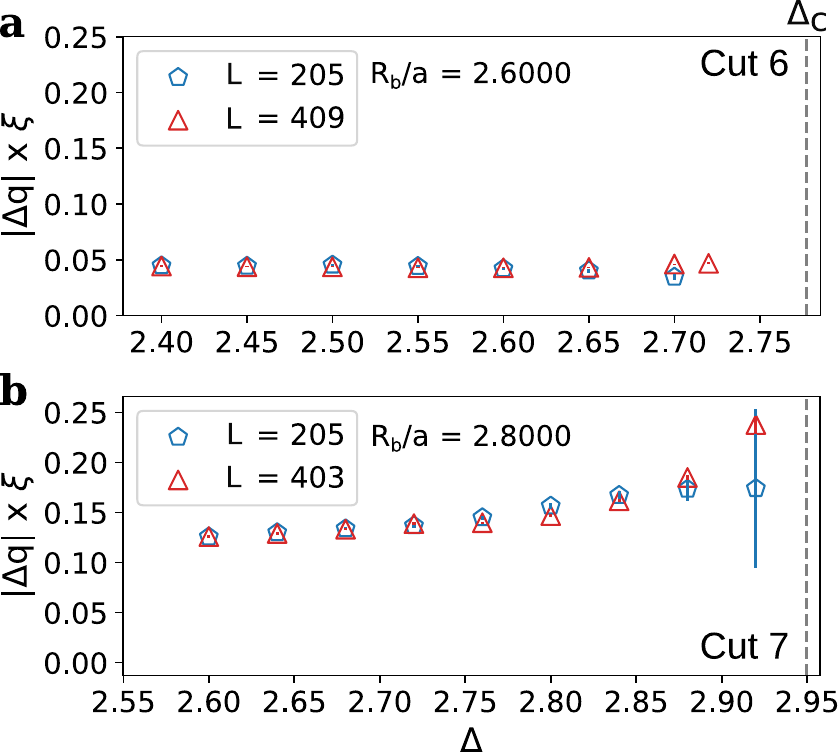}
    \caption{Scaling of the product $\Delta q \times \xi$ upon approaching (a) a chiral phase transition, and (b) a Kosterlitz-Talapov (KT) transition along Cuts 6 and 7 indicated in Fig.\ref{fig:z3z3z2sketch}. The correlation length $\xi$ was extracted from the density-density correlations of the operator $\hat s$. The location of the transitions are marked with vertical dashed lines. The error bars show the $95\%$ confidence interval for the point's location, incorporating fitting and error propagation.}
    \label{fig:z3Aq}
\end{figure}

To distinguish the chiral transition from the floating phase, we rely on the Huse-Fisher criteria: $\Delta q \times \xi$ remains finite upon approaching the chiral transition (see Fig.\ref{fig:z3Aq}(a)), but diverges when approaching the Kosterlitz-Thouless transition, as shown in Fig.\ref{fig:z3Aq}(b). This approach allows one to detect the floating phase even when its width is too small to study the scaling inside the floating phase.

\subsection{Ising transition between $\mathbb{Z}_3$ and $\mathbb{Z}_3\otimes \tilde{\mathbb{Z}}_2$ phases}
Similar to the previous case, the symmetry between the two legs of the ladder is spontaneously broken, which leads to an Ising transition separating $\mathbb{Z}_3$ and $\mathbb{Z}_3\otimes \tilde{\mathbb{Z}}_2$ phases, as indicated in Fig.\ref{fig:z3z3z2sketch}.
In Fig.\ref{fig:z3z3z2criticalexponents}, we provide numerical evidence of the Ising transition, including the numerically extracted scaling dimension of the order parameter  $m=\max(\hat a_i, \hat a_{i+1})$, central charge, and correlation length critical exponent $\nu$. All numerical values are in excellent agreement with CFT predictions\cite{difrancesco} for the Ising critical theory.

\begin{widetext}
   \centering
   
\begin{figure}[h] 
    \includegraphics[width=0.8\textwidth]{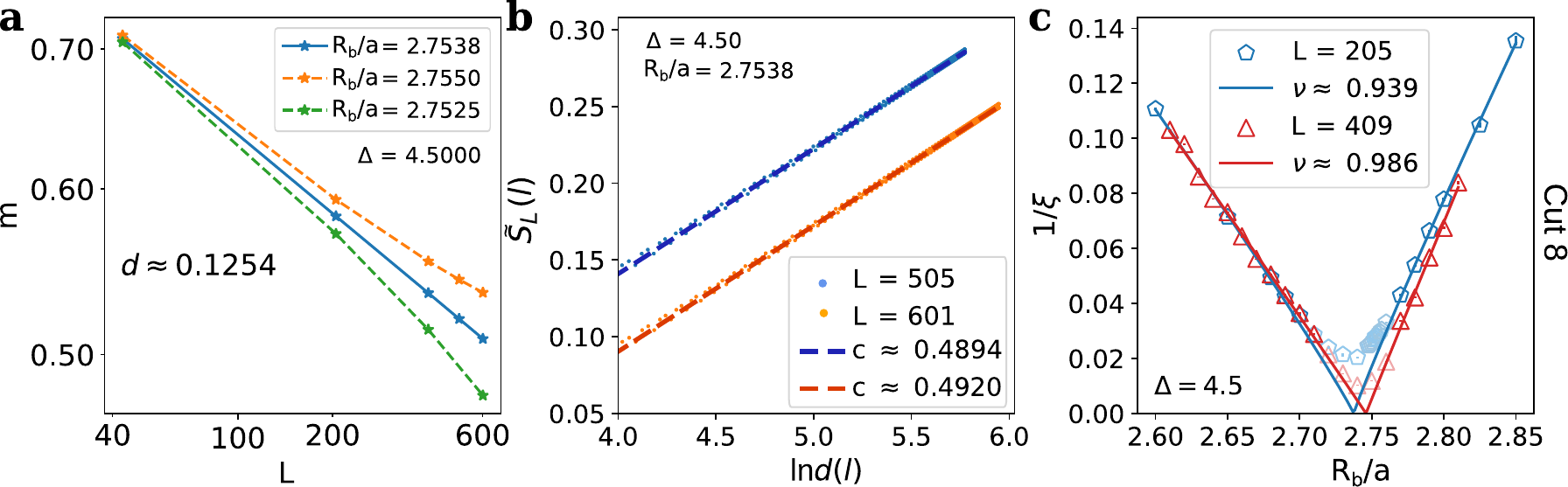}
    \caption{ Numerical evidences of the Ising transition between the $\mathbb{Z}_3$ and $\mathbb{Z}_3\otimes \tilde{\mathbb{Z}}_2$ phases. (a) Finite-size scaling of the order parameter $m=\max(\hat a_i, \hat a_{i+1})$. The extracted scaling dimension $d$ is in excellent agreement with CFT prediction $d=1/8$. (b) Scaling of the entanglement entropy with the conformal distance. Results for $L = 601$ are shifted by $-0.05$ for better visualization. The slope is in excellent agreement with the Ising central charge $c=1/2$. (c) Scaling of the inverse of the correlation length $1/\xi$ in the vicinity of the transition extracted from the density-density correlations of the operator $\hat s$. The extracted critical exponent $\nu$ agrees within $2\%$ of the CFT value $\nu=1$. Points that were not included in the fit are shown in pale.}
    \label{fig:z3z3z2criticalexponents}
\end{figure}
\end{widetext}

\subsection{Transition between the disordered and $\mathbb{Z}_3\otimes \tilde{\mathbb{Z}}_2$ phases}

\begin{figure}[h]
    \centering
    \includegraphics[width=\linewidth]{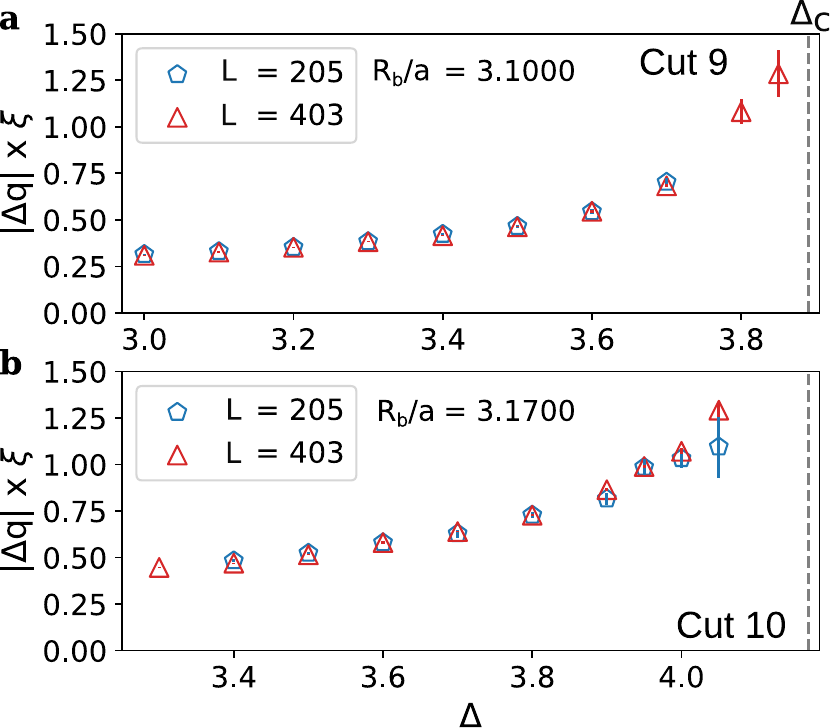}
    \caption{Scaling of the product $\Delta q \times \xi$ upon approaching (a) and (b) two Kosterlitz-Thouless transitions. The correlation length $\xi$ was extracted from the density-density correlations of the operator $\hat s$. The scaling is performed along the horizontal cuts, Cut 9 and Cut 10,  indicated in Fig.\ref{fig:z3z3z2sketch}. The error bars represent the $95\%$ confidence interval for the point's location, incorporating fitting and error propagation.}
    \label{fig:z3z2AqXi}
\end{figure}

The results for $\Delta q \times \xi$ presented in Fig.\ref{fig:z3z2AqXi} suggest that the transition between the disordered phase and the $\mathbb{Z}_3\otimes \tilde{\mathbb{Z}}_2$ ordered phase is always mediated by a floating phase. 

There are several possible scenarios, sketched in Fig.\ref{fig:multicritical}, on how the floating phase coating the $\mathbb{Z}_3$ transition might extend and encompass the $\mathbb{Z}_3\otimes\tilde{\mathbb{Z}}_2$ phase. In the simplest case, as depicted in Fig.\ref{fig:multicritical}(a), the Ising and two Pokrovsky-Talapov transitions meet at a multicritical point. In this case, both the nature of this multicritical point and the renormalization group flows are not clear. In a more plausible scenario, the Ising transition continues through the floating phase and separates two Luttinger liquid phases, one of which has spontaneously broken $\tilde{\mathbb{Z}}_2$ symmetry, as sketched in Fig.\ref{fig:multicritical}(b). In this configuration, the Ising and Pokrovsky-Talapov transitions are expected to be independent, with the crossing point exhibiting either Ising or Pokrovsky-Talapov criticalities, depending on the measured order parameter. A similar scenario has been recently reported in the context of interacting Majorana fermions\cite{10.21468/SciPostPhys.14.6.152}, although in that case, the Ising transition exited the floating phase via a Kosterlitz-Thouless transition. A third scenario, shown in Fig.\ref{fig:multicritical}(c), is very similar to the previous case, but places the Ising transition higher in the energy spectrum, and therefore, not necessary present in the Luttinger liquid phase in the ground-state phase diagram. Finally, if both $\mathbb{Z}_3$ and $\tilde{\mathbb{Z}}_2$ symmetries are broken simultaneously, the most natural expectation is a first order transition out of the $\mathbb{Z}_3\otimes\tilde{\mathbb{Z}}_2$ phase. However, in this scenario, it is not clear how the floating phase, which gains a finite width by the time it reaches the Ising transition, terminates.
We consider the scenario sketched in Fig.\ref{fig:multicritical}(b) to be the most plausible. However, the extremely narrow extent of the floating phase in our model prevents numerical verification. It would be worthwhile, nonetheless, to explore the aforementioned scenarios within simpler models, potentially with blockade constraints.

\begin{widetext}
   \centering
   
   \begin{figure}[h]
    \includegraphics[width=1\textwidth]{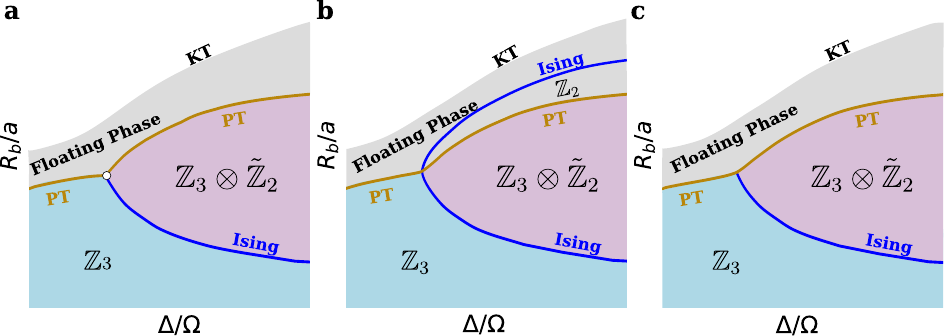}
    \caption{Sketch of the three possible scenarios for the multicritical point where the floating, $\mathbb{Z}_3$ and $\mathbb{Z}_3\otimes \mathbb{\tilde Z}_2$ phases meet. (a) The Ising and Pokrovsky-Talapov lines meet at a multicritical point, (b) The Ising critical line extends inside the floating phase as a higher-symmetry line separating two regions of the floating phase with spontaneously broken $\tilde{\mathbb{Z}}_2$ symmetry in one of them, (c) the Ising transition takes place higher in the energy spectrum after crossing the Pokrovsky-Talapov transition.}
    \label{fig:multicritical}
\end{figure}
\end{widetext}

\section{Discussion}
\label{sec:discussion}

To conclude, let us briefly compare our findings with
previous literature on Rydberg chains and ladders. Similar to a 1D array of atoms, the phase diagram of the Rydberg ladder model consists of lobes of crystalline gaped phases surrounded by a disordered phase, a large fraction of which is incommensurate. 
Interestingly, except for the standing alone checkerboard lobe, all crystalline phases appear in pairs. These paired phases share the same periodicity in the rung density, but are distinguished by spontaneously broken symmetry between the two legs in one of the phases.  This asymmetry gives rise to an Ising transition between each pair of ordered phases.
The bi-lobed structure of the gaped phases has been overlooked in the previous experimental and numerical study of the Rydberg ladder\cite{zhang2024probing}. It is possible, however, that larger inter-chain distance considered in Ref.\onlinecite{zhang2024probing} might affect the extent of each phase.

In contrast, the resonant phases ---gaped phases preserving the symmetry between the two legs of the ladder---have been reported in the recent analytical study of a Rydberg blockade model on a ladder\cite{eck2023critical}. However, the phase diagram by Eck and Fendley\cite{eck2023critical} featured two different resonant $\mathbb{Z}_3$ phases: a phase with a symmetric rung state $\frac{1}{\sqrt{2}}\left(\ket{10} + \ket{01} \right)$, called $\mathbb{Z}_3^+$, and one with antisymmetric rungs  $\frac{1}{\sqrt{2}}\left(\ket{10} - \ket{01} \right)$,  labeled $\mathbb{Z}_3^-$. The $\mathbb{Z}_3$ phases in the present paper corresponds to a symmetric $\mathbb{Z}_3^+$ one in Eck and Fendley's notation. 
By contrast, its antisymmetric counterpart appears due to the hopping term  $d_{2i}d^\dagger_{2i+1}+\mathrm{h.c.}$ introduced in Ref.\onlinecite{eck2023critical} to keep the model integrable, but absent in the experimentally feasible model (\ref{eq:ham}) that we consider here.

The complex bi-lobbed structure of the density-wave phases also leads to exotic critical behavior at their boundaries. In particular, we observed a fusion of two Ising transitions, resulting in an extended region with a conformal Ashkin-Teller transition, followed by a $\mathbb{Z}_4$-chiral transition. This suggests that, for some reason, the two symmetries $\mathbb{Z}_2$ and $\mathbb{\tilde Z}_2$ are simultaneously broken at the upper boundary of $\mathbb{Z}_2\otimes \tilde{\mathbb{Z}}_2$ phase.
By contrast to the model proposed by Eck and Fendley\cite{eck2023critical}, where an extended Ashkin-Teller criticality has been preserved by the integrability of the model, in the present case, the appearance of the finite interval of the conformal Ashkin-Teller transition is very surprising, as the model we consider is non-intergrable.
In our model, the conformal criticality is preserved by the lack of incommensurability surrounding the multicritical point where two Ising transitions fuse.  
Recently, the fusion of two Ising transition leading to a single Ashkin-Teller critical point immediately followed by the chiral transition has been predicted in the multi-component Rydberg array\cite{chepiga2024tunable}. In that case, the disorder line separating a commensurate and incommensurate areas of the disordered phase hits the boundary of the crystalline phase at the multi-critical point. In the present case, this happens at a different point leaving a finite-interval with the  Ashkin-Teller transition  uprerturbed by the chiral perturbations.



Another interesting opportunity is raised from the connection with multi-component Rydberg atoms, where the nature of the Ashkin-Teller criticality as well as the extent of the chiral transition following it have been tuned continuously by the relative ratio of two Rabi frequencies\cite{chepiga2024tunable}.
In the two leg-ladder considered here, we observe a similar enhancement of symmetries, leading to a similar fusion of two Ising transitions. It is therefore natural to expect that a similar level of control over phases and phase transitions can be achieved by tuning the ratio of inter-atomic spacing within each chain and between them. It would be noteworthy to explore how the phase diagram in Fig.\ref{fig:phasediagram} and the nature of the quantum phase transitions change with variations in the relative height of the ladder that can easily be controlled in experiment. 

Furthermore, our study underscores the importance of the appropriate choice of the order parameter and operators in numerical and experimental studies of quantum phase transitions. In Fig.\ref{fig:fsscuts}, we demonstrated that the scaling dimension along the Ashkin-Teller ($c=1$) transition interval behaves differently for different operators, reflecting the complex nature of this critical line. For a given operator, its scaling dimension is related to its correlation length critical exponent. 
Since the nature of quantum phase transitions in Rydberg arrays is traditionally investigated using the quantum Kibble-Zurek mechanism, selecting the appropriate correlation function and kink operators\cite{zurek2005dynamics, dziarmaga2005dynamics} is a key for experimental observation of this exotic transition.

Finally, there is another open question raised by our study---the nature of the multicritical point where the $\tilde{\mathbb{Z}}_2$ Ising transition comes across the floating phase into $\mathbb{Z}_3$ phase. In Fig.\ref{fig:multicritical}, we already listed and briefly described several possible scenarios. We highlighted that the most plausible scenario is one where the Ising transition continues through the floating phase (see Fig.\ref{fig:multicritical}(b)), which appears due to an emergent U(1) symmetry\cite{verresen2019stable}. If our conjecture is correct, one can expect an emergent  $\mathcal{N} = (1, 1)$ supersymmetry \cite{FODA1988611,PhysRevLett.102.176404} along the line of Ising transition superposed with the  U(1) symmetric Luttinger liquid phase. Due to the extremely narrow width of the floating phase and overall large complexity of calculations of the model with van der Waals potential, this problem cannot be resolved numerically within the current framework. We believe that the most promising approach would be to construct a related constrained model that allows for more accurate numerical simulations and, ideally, analytically solvable at least in certain limits or along the lines with higher symmetry. While this approach lies outside the immediate scope of our research, it represents an intriguing avenue for future investigation.

 \section{Acknowledgments}
We thank Paul Fendley and Luisa Eck for insightful comments. NC acknowledges useful discussions with Andreas L\"auchli and Frederic Mila.
  This research has been supported by Delft Technology Fellowship.  
  Numerical simulations have been performed at the DelftBlue HPC and at the Dutch national e-infrastructure with the support of the SURF Cooperative.
  
\appendix

\section{Beyond $\mathbb{Z}_3 \otimes \mathbb{\tilde Z}_2$}
\label{sec:z4}
\begin{figure}[h]
    \centering
    \includegraphics[width=\linewidth]{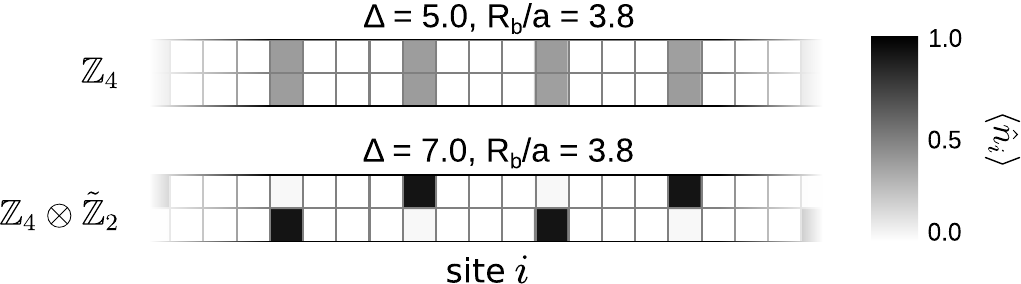}
    \caption{Density profiles for other gaped phases not discussed in the main text. Specifically, the $\mathbb Z_4$ phase was obtained for $\Omega = 1$, $\Delta = 5.0$ and $R_b/a = 3.8$, while the $\mathbb Z_4\otimes\mathbb{\tilde Z}_2$ phase was obtained for $\Omega = 1$, $\Delta = 7.0$ and $R_b/a = 3.8$.  Each square corresponds to a lattice site, with its color indicating the local density according to the color bar. We show profiles  only for the central part of a long chain.}
    \label{fig:beyond}
\end{figure}

There are other crystalline phases above $\mathbb Z_3 \otimes \mathbb{\tilde Z}_2$ that have not been studied in detail due to extremely slow convergence. In Fig.\ref{fig:beyond} we present examples of the $\mathbb{Z}_4$ and $\mathbb{Z}_4\otimes\tilde{\mathbb{Z}}_2$ phases, continuing the sequence of the bi-lobed structures.

\section{Influence of the operator choice on the extracted correlation length $\xi$ and wave vector $q$}
\label{sect:operatorcorrelation}
As discussed in the main body of this article, different operators exhibit distinct scaling behaviors near the critical points. These variations arise because each operator captures unique aspects of the broken symmetries associated to the phase transitions. In this section, we illustrate this phenomenon by comparing the scaling of the inverse correlation length $1/\xi$ extracted from the density-density correlations for the three operators used in our calculations---$\hat t$, $\hat s$ and $\hat a$---as defined in Eq.\ref{eq:operator}.

\begin{figure}[h]
    \centering
    \includegraphics[width=\columnwidth]{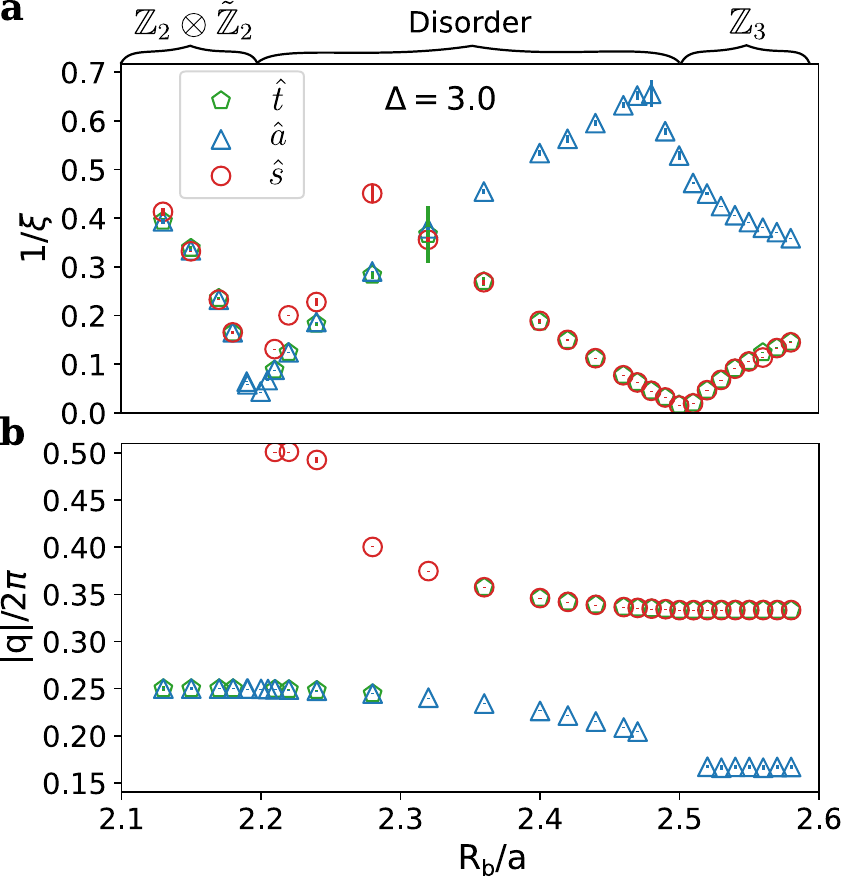}
    \caption{Comparison of (a) the inverse of the correlation length $1/\xi$ and (b) the wave-vector $q$ calculated from the density-density correlation of the operators $\hat t$, $\hat a$ and $\hat s$ for a vertical cut comprising phases $\mathbb{Z} _2 \otimes \mathbb{\tilde Z}_2$, $\mathbb Z_3$ and the disordered phase in between. The expressions for the operators $\hat t$, $\hat a$ and $\hat s$ are described in Eq.\ref{eq:operator}. Error bars correspond to the $95\%$ confidence interval for the fit, estimated as $\pm 1.96$ times the standard error of the predicted values, obtained from the covariance matrix of the fit parameters.}
    \label{fig:z2z2z3xiqoperator}
\end{figure}

\begin{figure*}[t]
    \centering
    \includegraphics[width=0.8\linewidth]{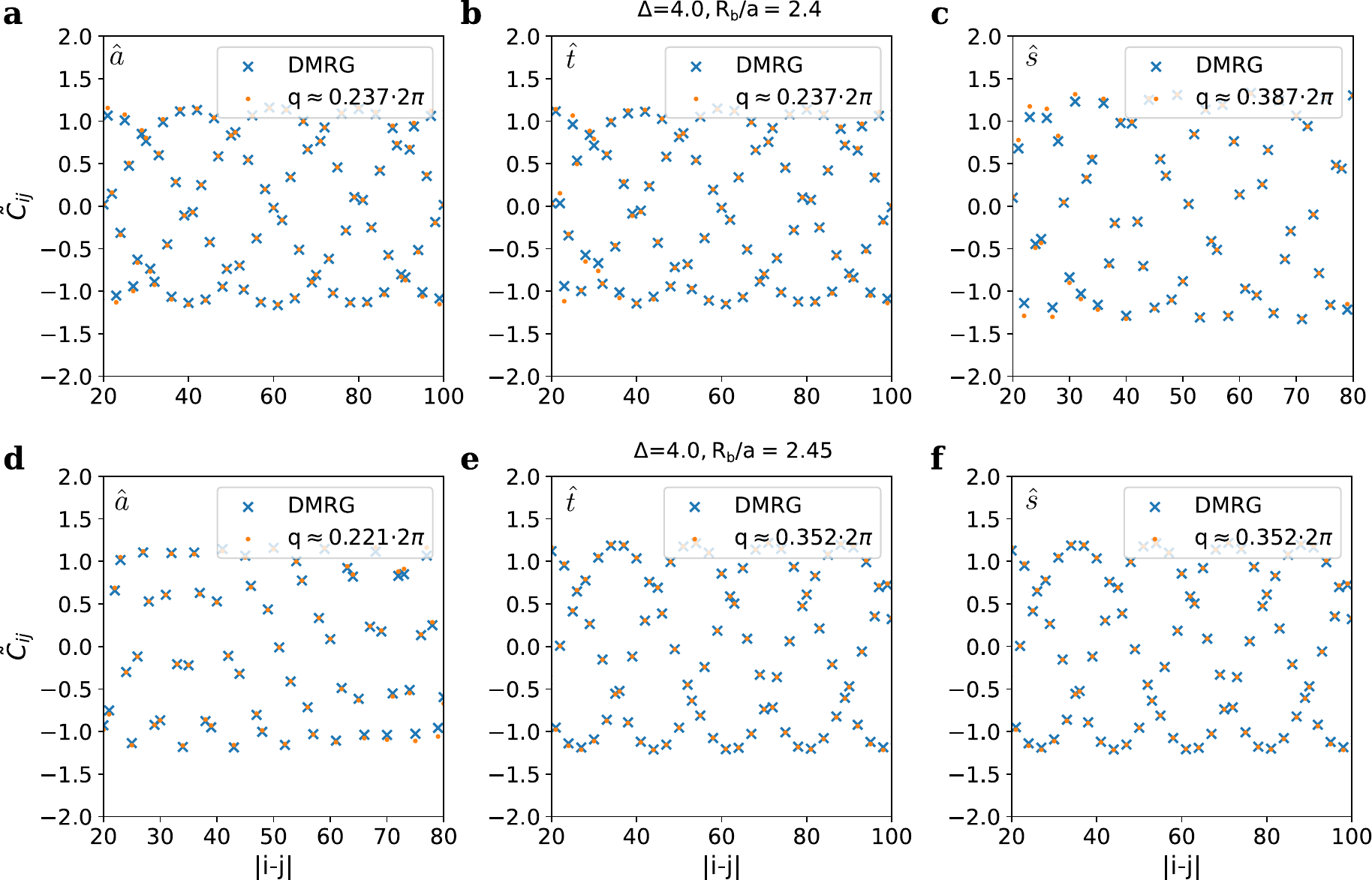}
    \caption{Fitting of the wave-vector $q$ to the oscillations of the correlation function, calculated for three distinct operators: $\hat{a}$, $\hat{t}$, and $\hat{s}$ defined in Eq.\ref{eq:operator}, in two different scenarios. (a) The operators $\hat a$ and $\hat{t}$ yield the same value for $q$, which differs from that obtained using $\hat{s}$. (b) The operators $\hat{s}$ and $\hat{t}$ produce the same value for $q$, distinct from the result obtained using $\hat a$. The definitions of the operators $\hat a$, $\hat{t}$, and $\hat{s}$ are provided in Eq.\ref{eq:operator}.}
    \label{fig:z2z2z3qoperator}
\end{figure*}
Fig. \ref{fig:z2z2z3xiqoperator}(a) presents the scaling of $1/\xi$ for a vertical cut comprising the phases $\mathbb{Z}_2\otimes\mathbb{\tilde Z}_2$, $\mathbb{Z}_3$ and the disordered phase in between. Within the $\mathbb{Z}_2\otimes\mathbb{\tilde Z}_2$, phase the correlation length is independent of the operator used. However, in the disordered phase, the correlation length associated with $\hat s$ is slightly smaller than that of $\hat a$ and $\hat t$, but our results are consistent with all correlation lengths diverging  at the same critical point $R_b/a$, compatible with the direct transition between the disordered and $\mathbb{Z}_2\otimes\mathbb{\tilde Z}_2$ ordered phases.

The situation is drastically different upon entering the  $\mathbb{Z}_3$ phase: the correlation lengths associated with $\hat s$ and $\hat t$, both susceptible to spontaneously broken symmetry in the $\mathbb{Z}_3$ phase, diverge around $R_b/a\approx 2.5$. In contrast, the correlation length associated with $\hat a$, which tracks states with broken $\mathbb{\tilde Z}_2$ symmetry, remains finite across the transition. Within the $\mathbb{Z}_3$ phase, the correlation lengths calculated with $\hat t$ and $\hat s$ decrease with the distance to the transition, while the one extracted from $\hat a$ increases as we move out of the transition. 

The behavior of $\xi$ seems to be closely related to that of the wave-vector $q$, as illustrated in Fig.\ref{fig:z2z2z3xiqoperator}(b). Within the $\mathbb{Z}_2 \otimes \mathbb{\tilde{Z}}_2$ phase, the values of $q$ obtained from operators $\hat t$ and $\hat a$ are both $q = 2\pi/4$, which differ from $q = 2\pi /2$ extracted from $\hat s$.

\begin{figure}[h]
    \centering
    \includegraphics[width=\linewidth]{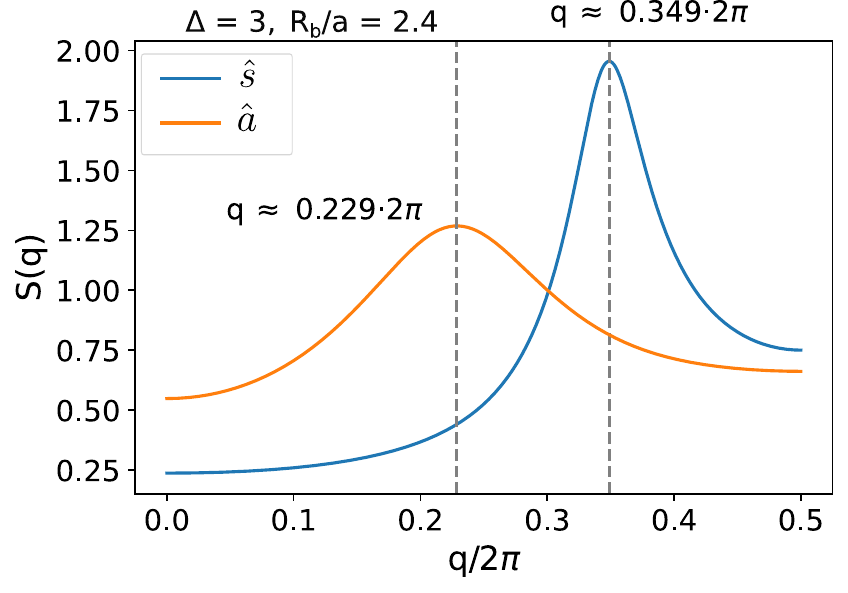}
    \caption{Extraction of the wave-vector $q$ from the peak of the two-particle correlation structure factor $S(q)$, computed for two different operators, $\hat{s}$ and $\hat a$, at a specific point within the disordered phase between the $\mathbb{Z}_2 \otimes \mathbb{Z}_2$ and $\mathbb{Z}_3$ phases. The operator $\hat{s}$ yields a value of $q$ close to $2\pi/3$, corresponding to the occupancy level of the $\mathbb{Z}_3$ phase, while the operator $\hat a$ produces a value of $q$ near $2\pi/4$, which aligns with the occupancy level of the $\mathbb{Z}_2 \otimes \mathbb{Z}_2$ phase.}
    \label{fig:structurefactor} 
\end{figure}

In the disordered phase, $q$ derived from $\hat a$ transitions from $q = 2\pi /4$ in the vicinity of the  $\mathbb{Z}_2 \otimes \mathbb{\tilde{Z}}_2$ phase to $q = 2\pi /6$, at the $Z_3$ phase, while $q$ goes from $q = 2\pi /2$ near the $\mathbb{Z}_2 \otimes \mathbb{\tilde{Z}}_2$ phase to $q = 2\pi /3$ at the $\mathbb{Z}_3$ phase.  
The behavior of $q$ extracted from $\hat t$ is more complex: it matches that of $\hat a$ near the $\mathbb{Z}_2 \otimes \mathbb{\tilde{Z}}_2$ phase and aligns with $\hat s$ closer to the $\mathbb{Z}_3$ phase. A sudden change in $q$ for $\hat t$ occurs at the point where the values of $\xi$ calculated for all three operators coincide ($R_b/a\approx 2.32$). 
An example of the extraction of the $q$ vector for the case when $\hat t$ coincides with $\hat s$ and when it coincides with $\hat a$ is shown in Fig.\ref{fig:z2z2z3qoperator}.
As shown in Fig.\ref{fig:structurefactor}, this effect also occurs when $q$ is extracted from the structure factor.

\begin{figure}[h]
    \centering
    \includegraphics[width=\linewidth]{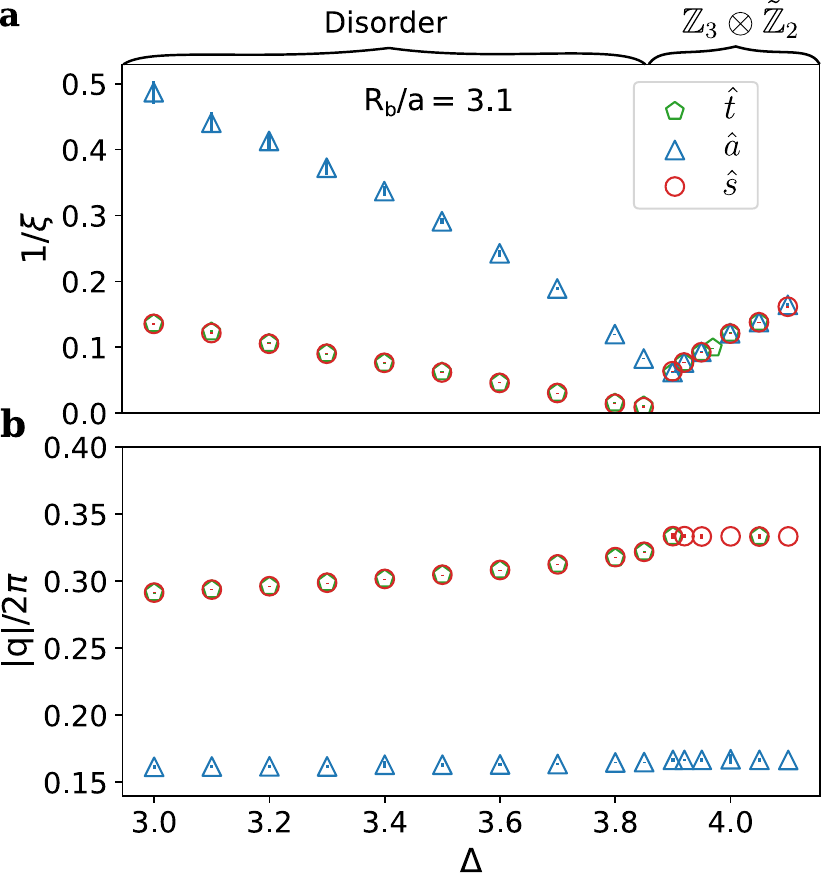}
     \caption{Comparison of (a) the inverse of the correlation length $1/\xi$ and (b) the wave-vector $q$ calculated from the density-density correlations of operators $\hat t$, $\hat a$ and $\hat s$ defined in Eq.\ref{eq:operator}, for an horizontal cut from the disordered to the $\mathbb{Z}_3 \otimes \mathbb{\tilde Z}_2$ phase. The error bars depict the $95\%$ confidence region for the values, estimated using $\pm1.96$ times the standard error of the predicted values}
    \label{fig:z3z2operator}
\end{figure}

One would naively expect that the relative behavior between the three aforementioned operators observed near the $\mathbb{Z}_2\otimes\tilde{\mathbb{Z}}_2$ phase would also be observed near the $\mathbb{Z}_3\otimes\tilde{\mathbb{Z}}_2$ phase. However, this is not the case. As illustrated in Fig\ref{fig:z3z2operator}(a), within the  $\mathbb{Z}_3\otimes\tilde{\mathbb{Z}}_2$ phase, the three operators give the same $\xi$. However, in contrast to the vicinity of the $\mathbb{Z}_2\otimes\tilde{\mathbb{Z}}_2$ phase, in the disordered phase close to the  $\mathbb{Z}_3\otimes\tilde{\mathbb{Z}}_2$ phase, the correlation length associated with $\hat{t}$ and $\hat{s}$ are similar and compatible with the floating phase (approaching the transition with zero slope) and distinct from operator $\hat{a}$. This scenario suggests that the floating phase responsible for the spontaneously broken $\mathbb{Z}_3$ symmetry is independent of the transition breaking $\tilde{\mathbb{Z}}_2$. In our numerical simulations, we cannot zoom closely enough to resolve the nature of the phase transition breaking $\tilde{\mathbb{Z}}_2$; however, it could be be either a first order transition or an Ising transition separating the two Luttinger liquids, as sketched in Fig.\ref{fig:multicritical}(b).
Wave-vector $q$ presented in Fig.\ref{fig:z3z2operator}(b) and extracted with $\hat{t}$ and $\hat{s}$ is compatible with commensurate-incommensurate transition (including the floating phase scenario).

\section{Central charge for the transition out of $\mathbb{Z}_2\otimes\mathbb{\tilde Z}_2$}
\label{sec:centralcharge}
\begin{figure}[h]
    \centering
    \includegraphics[width=0.9\columnwidth]{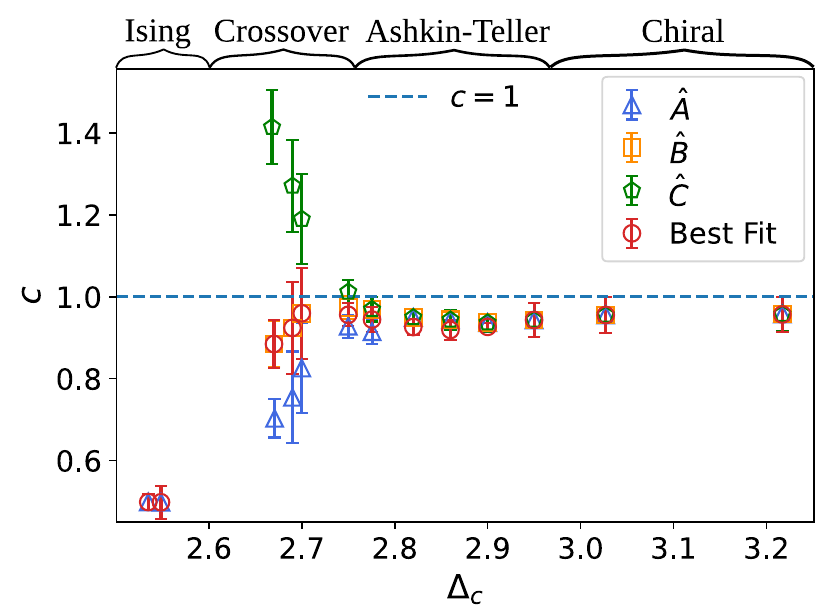}
    \caption{Comparison of central charge $c$ for various cuts along the upper critical line of the transition out of $\mathbb{Z}_2 \otimes \mathbb{\tilde Z}_2$, using critical point estimated from the scaling of the order parameters $\hat A$, $\hat B$ and $\hat C$, as defined in Eq.\ref{eq:fssoperator}. These results are compared to the best-fit value of $c$, defined as the value of $c$ that shows minimal variation with the system size. Error bars correspond to the $95\%$ confidence interval for the fit, estimated as $\pm 1.96$ times the standard error of the predicted values, obtained from the covariance matrix of the fit parameters.}
    \label{fig:z2z2ccuts}
\end{figure}

In this section, we compute the central charge $c$ along the critical line of the transition out of $\mathbb{Z}_2\otimes\tilde{\mathbb{Z}}_2$.
The location of the transition generally depends on the choice of the operator---$\hat A$, $\hat B$ and $\hat C$---that we use to perform the finite-size scaling described in Eq.\ref{eq:fssoperator}. We compare the central charge extracted at the critical points associates with each of the three operators with the best-fit value for $c$, determined as the value of $c$ that shows minimal variation with the system size. Fig \ref{fig:z2z2ccuts} summarizes our main results.

This figure reveals 4 distinct regimes, which coincide with the regimes observed for the finite-size scaling of the order parameter in Fig.\ref{fig:fsscuts}(a):
\begin{itemize}
    \item $\Delta_c < 2.6$. This range corresponds to the transition from $\mathbb{Z}_2$ to $\mathbb{Z}_2 \otimes \mathbb{\tilde Z}_2$. Here, only operator $\hat A$ effectively distinguishes the two phases. The central charge obtained with $\hat A$ aligns with the best-fit value, $c \approx 0.5$, consistent with CFT predictions for the Ising universality class.
    \item $2.6 < \Delta_c < 2.75$. This region, located close to the critical point, appears to exhibit a crossover behavior between the Ising and Ashkin-Teller transitions. The location of the critical points differs slightly depending on the order parameter, which results in variations in the value of $c$. In this range, operator $\hat B$ provides the closest match to the best-fit value of $c$. For this operator, $c$ shows an intermediate value between the $c = 0.5$ predicted for the Ising transition and $c = 1$ predicted for the Ashkin-Teller transition. As we move further from the multicritical point into the Ashkin-Teller region, $c$ approaches 1.
    \item $2.75 < \Delta_c < 2.95$. This interval corresponds to the Ashkin-Teller region. Conformal field theory predicts $c = 1$ for the Ashkin-Teller universality class. In this region, the choice of operator has minimal impact on the location of the critical point, and the computed values of $c$ are nearly identical across all operators. However, the best-fit $c$ is slightly lower than the values obtained from the operators, although all measurements remain close to 1, within $5\%$ relative error. The small deviations from the theory prediction $c=1$ are likely due to finite size effects and are comparable with previously reported values for the Ashkin-Teller multicritical point in a related models\cite{chepiga2021kibble}.
    \item $2.95 < \Delta_c$. Chiral transition. All operators identify the critical point at the same location, and the obtained effective central charge $c$ is equal to the best-fit $c$. Chiral transition is not conformal, and the central charge is not defined.
\end{itemize}

\bibliographystyle{unsrt}
\bibliography{main_arxiv}
\pagebreak
\end{document}